\global\def\draftcontrol{0}

   \def\versionno{ Time-Dependent Processes at finite coupling -- draft   }

\catcode`\@=11

\expandafter\ifx\csname draftcontrol\endcsname\relax\global\def\draftcontrol{0}
\fi

{\count255=\time\divide\count255 by 60
\xdef\hourmin{\number\count255}
\multiply\count255 by-60\advance\count255 by\time
\xdef\hourmin{\hourmin:\ifnum\count255<10 0\fi\the\count255}}
\def\draftdate{\number\month/\number\day/\number\year\ \ \ \hourmin }

\newcommand\makepapertitle{\par
  \begingroup
    \renewcommand\thefootnote{\@fnsymbol\c@footnote}%
    \def\@makefnmark{\rlap{\@textsuperscript{\normalfont\@thefnmark}}}%
    \long\def\@makefntext##1{\parindent 1em\noindent
            \hb@xt@1.8em{%
                \hss\@textsuperscript{\normalfont\@thefnmark}}##1}%
     \newpage
     \global\@topnum\z@   
     \@makepapertitle
     \thispagestyle{empty}\@thanks
  \endgroup
  \setcounter{footnote}{0}%
  \global\let\thanks\relax
  \global\let\makepapertitle\relax
  \global\let\@makepapertitle\relax
  \global\let\@thanks\@empty
  \global\let\@author\@empty
  \global\let\@date\@empty
  \global\let\@title\@empty
  \global\let\title\relax
  \global\let\author\relax
  \global\let\date\relax
  \global\let\and\relax
  \def\version{\let\version\@version\@gobble}
}
\def\@makepapertitle{%
  \newpage
   \ifnum\draftcontrol=1 {}
   \version\versionno
   \vskip 3em%
   \else
   \hfill\hbox to 3cm {\parbox{4cm}{\@pubnum}\hss}%
   \vskip 3em%
   \fi
   \begin{center}%
   \let \footnote \thanks
     {\LARGE {\@title}}%
     \vskip 1.5em%
     {\normalsize
       \lineskip .5em%
       \begin{tabular}[t]{c}%
         \@author
       \end{tabular}\par}%
     \vskip 1.5em%
     {\@bstract}%
     \end{center}%
     \vskip 1.5em
     \@date%
   \par
}

\gdef\@pubnum{}
\def\pubnum#1{%
  \gdef\@pubnum{#1}}

\gdef\@bstract{}
\def\Abstract#1{%
  \gdef\@bstract{%
   \parbox{\textwidth-0pc}{%
   \centerline{\bf Abstract}\penalty1000%
\kern.2cm%
\noindent
\renewcommand\baselinestretch{1.0}%
{#1}}}
}

\def\ps@paper{\let\@mkboth\@gobbletwo%
     \ifnum\draftcontrol=1
    \def\@oddfoot{\hbox to \textwidth{\tiny \versionno \hfil\tiny\draftdate}%
    \hskip -\textwidth \hbox to \textwidth{\hfil\rm\thepage\hfil}}%
     \else\def\@oddfoot{\hbox to \textwidth{\hfil\rm\thepage\hfil}}
     \fi
     \let\@evenfoot\@oddfoot
}

\def\body{\clearpage
          \pagestyle{paper}
    }

\def\@version#1{\ifnum\draftcontrol=1
\typeout{}\typeout{#1}\typeout{}
\vskip3mm\centerline{\hbox{\fbox{\normalsize{\tt DRAFT -- #1 -- }
                   {\draftdate}}}}\vskip3mm
\fi}
\let\version\@version
\long\def\eqlabel#1{\ifnum\draftcontrol=1
                    \tag@false  
                    \tag*{(\theequation) \hbox to -0.2cm{\hspace{0cm}\small{#1}\hss}}
                    \refstepcounter{equation}
                    \edef\@currentlabel{\theequation}
                    \ltx@label{#1}          
                    \else
                    \label{#1}
                    \fi
                    }
\let\st@bibitem\@bibitem
\let\st@lbibitem\@lbibitem
\ifnum\draftcontrol=1
  \def\@bibitem#1{%
    \st@bibitem{#1}\a@@label{#1}\ignorespaces}
  \def\@lbibitem[#1]#2{%
    \st@lbibitem[#1]{#2}\a@@label{#2}\ignorespaces}
  \def\a@@label#1{%
    \gdef\a@lab{\smash{\normalfont\small#1}}
    \ifvmode
      \if@inlabel
        \global\setbox\@labels\hbox{%
          \llap{\a@lab\let\a@lab\relax
                \kern\@totalleftmargin\kern\marginparsep}%
          \box\@labels}%
      \fi
    \fi}
\fi

\documentclass[12pt,letterpaper]{article}

\usepackage{amsmath,amssymb,array,calc,epsfig}

\ifnum\draftcontrol=1
\fi

\renewcommand\baselinestretch{1.25}
\renewcommand\section{\@startsection {section}{1}{\z@}%
                                   {-3.5ex \@plus -1ex \@minus -.2ex}%
                                   {2.3ex \@plus.2ex}%
                                   {\normalfont\large\bfseries}}
\renewcommand\subsection{\@startsection{subsection}{2}{\z@}%
                                   {-3.25ex\@plus -1ex \@minus -.2ex}%
                                   {1.5ex \@plus .2ex}%
                                   {\normalfont\normalsize\bfseries}}
\renewcommand\subsubsection{\@startsection{subsubsection}{3}{\z@}%
                                   {-3.25ex\@plus -1ex \@minus -.2ex}%
                                   {1.5ex \@plus .2ex}%
                                   {\normalfont\normalsize\it}}
\renewcommand\paragraph{\@startsection{paragraph}{4}{\z@}%
                                   {-3.25ex\@plus -1ex \@minus -.2ex}%
                                   {1.5ex \@plus .2ex}%
                                   {\normalfont\normalsize\bf}}

\numberwithin{equation}{section}

\def\ie{{\it i.e.}}

\def\revise#1       {\raisebox{-0em}{\rule{3pt}{1em}}%
                     \marginpar{\raisebox{.5em}{\vrule width3pt\
                     \vrule width0pt height 0pt depth0.5em
                     \hbox to 0cm{\hspace{0cm}{%
                     \parbox[t]{4em}{\raggedright\footnotesize{#1}}}\hss}}}}

\newcommand\nxt[1]  {\\\fnxt#1}

\def\calc         {{\cal C}}
\def\cald         {{\cal D}}
\def\cale         {{\cal E}}
\def\calf         {{\cal F}}

\def\cali         {{\cal I}}

\def\call         {{\cal L}}

\def\caln         {{\cal N}}
\def\calo         {{\cal O}}

\def\calr         {{\cal R}}
\def\cals         {{\cal S}}

\def\sqr#1#2{{\vcenter{\vbox{\hrule height.#2pt
 \hbox{\vrule width.#2pt height#1pt \kern#1pt
 \vrule width.#2pt}\hrule height.#2pt}}}}

\def\a{\alpha}
\def\b{\beta}
\def\w{\omega}
\def\r{\rho}
\def\dd{\delta}
\def\e{\epsilon}

\def\ga{\gamma}

\def\aa1{\phi}
\def\cc1{\psi}

\def\t{\tau}

\def\l{\lambda}
\def\ha{\hat{a}}
\def\hb{\hat{b}}
\def\hc{\hat{c}}
\def\hal{\hat{\alpha}}
\def\hp{\hat{\phi}}
\def\he{\hat{\eta}}

\newcommand{\eq}{\begin{equation}}
\newcommand{\eqx}{\end{equation}}
\newcommand{\eqn}{\begin{eqnarray}}
\newcommand{\eqnx}{\end{eqnarray}}

\catcode`\@=12

\begin{document}


\title{\bf Shear viscosity of boost invariant plasma at finite coupling}
\pubnum{%
UWO-TH-08/2}

\date{January 2008}

\author{
Alex Buchel\\[0.4cm]
\it Department of Applied Mathematics\\
\it University of Western Ontario\\
\it London, Ontario N6A 5B7, Canada\\[0.2cm]
\it Perimeter Institute for Theoretical Physics\\
\it Waterloo, Ontario N2J 2W9, Canada
}

\Abstract{
We discuss string theory $\a'$ corrections in the dual description of
the expanding boost invariant $\caln=4$ supersymmetric Yang-Mills
plasma at strong coupling. We compute finite 't Hooft coupling
corrections to the shear viscosity and find that it disagrees with the
equilibrium correlation function computations. We comment on the
possible source of the discrepancy.
}

\makepapertitle

\body

\version\versionno

\section{Introduction}
Recent work by Janik and collaborators \cite{j1,j2,j3} (see also \cite{j15}) opened a possibility to 
study non-stationary/non-equilibrium processed in gauge theories using their dual string theory 
formulation \cite{m1,m2}. In fact, one might try to use gauge theory/string theory correspondence 
to {\it define} non-equilibrium Quantum Field Theory dynamics\footnote{For an approach 
alternative to \cite{j1} see \cite{muk}.}. For example: in principle, one 
can use AdS/CFT correspondence to extend Muller-Israel-Stewart theory of dissipative relativistic hydrodynamics 
\cite{m,is} to arbitrary order in deviation from the equilibrium.  

Even though the framework proposed in \cite{j1} can not be consistently implemented 
within a supergravity approximation of the string theory dual to $\caln=4$ supersymmetric 
Yang-Mills (SYM) plasma in the Bjorken flow \cite{bj} (see \cite{bbhj}), if correctly reproduces 
the equilibrium and near-equilibrium properties of the $\caln=4$ SYM plasma obtained from the 
equilibrium correlation functions, \ie, the equilibrium equation of state \cite{ads1},
its shear viscosity at infinite 't Hooft coupling \cite{u2}, and its relaxation time\footnote{Related work appeared in 
\cite{sol1,sol2,sol3}.}
\cite{sl}. Furthermore, this agreement appears to be unaffected by the subtleties 
associated with the breakdown of the supergravity approximation. Given substantial 
potential of the Janik's framework\footnote{One can formulate and study questions such as 
dynamical hadronization in this approach.}, we believe it is important to further 
explore its regime of validity.

As we already mentioned, the shear viscosity of the $\caln=4$ plasma at infinite 
coupling is a robust prediction within Janik's framework \cite{j2}. 
In this paper we compute finite t' Hooft coupling corrections 
(corresponding to string theory $\a'$ corrections in the dual formulation)  
to the $\caln=4$ plasma. Since physically equivalent results extracted from the 
equilibrium correlation functions are already available \cite{sh1,sh2}, 
our analysis provides a test of the framework \cite{j1} beyond 
the supergravity approximation. We find that the finite coupling 
corrections to the shear viscosity of the $\caln=4$ plasma extracted from the 
string theory dual to its Bjorken flow disagrees with the equilibrium correlation 
function computations.

The paper is organized as follows. In the next section we review the results of \cite{sh1,sh2},
the dual string theory model within which such computation were performed, and emphasize the 
self-consistency of the obtained results\footnote{For a recent review of the Janik's framework and 
its clarification see \cite{bbhj}.}. We further summarize our new results for 
shear viscosity at finite 't Hooft coupling. In section 3 we describe our  string theory computational 
framework introduced in \cite{bal}. The analysis are incredibly complicated and thus we focus 
on explaining the consistency checks on our analysis (fortunately, there are lots of such consistency checks).
In section 4 we summarize equations of motion describing $\a'$ correction to the supergravity background dual to 
late-time Bjorken flow of the $\caln=4$ plasma. We also present analytic solutions to these equations. 
In section 5 we map the results of the string theory computation to the near-equilibrium Muller-Israel-Stewart (MIS)
theory 
and extract the leading correction to the SYM plasma shear viscosity. We conclude in section 6.
Appendix contains some technical details; further computation details are available from the author 
upon request.
 
\section{Shear viscosity of the $\caln=4$ SYM plasma at finite 't Hooft coupling }  
 
Consider $\caln=4$ $SU(n_c)$ SYM plasma in the ('t Hooft) large-$n_c$ limit:
we take the Yang-Mills coupling $g_{YM}^2\to 0$ and $n_c\to \infty$ in such a way 
that the 't Hooft coupling $\l$,
\begin{equation}
\l=g_{YM}^2 n_c\,,
\eqlabel{ldef}
\end{equation}
remains finite. We are interested in the transport properties of the plasma at strong coupling, \ie, 
$\lambda\gg 1$. In this limit the useful description of the $\caln=4$ plasma is in terms 
of holographically dual background of near-extremal black 3-brane geometry in the supergravity approximation 
of the type IIb string theory. The supergravity approximation is exact at $\l=\infty$, 
in which case one finds \cite{u2,pol1,pol2} the speed of sound $c_s$, the shear viscosity $\eta$,
and the bulk viscosity $\xi$ correspondingly
\begin{equation}
c_s=\frac{1}{\sqrt{3}}\,,\qquad \eta=\frac{\pi}{8} n_c^2 T^3\,,\qquad \xi=0\,.
\eqlabel{n4res}
\end{equation} 
In the supergravity approximation the entropy density $s$ is 
\begin{equation}
s=\frac{\pi^2}{2}n_c^2 T^3\,,
\end{equation}  
leading to the viscosity-to-entropy ratio 
\begin{equation}
\frac{\eta}{s}=\frac{1}{4\pi}\,.
\eqlabel{etas}
\end{equation}  

In a hydrodynamic approximation to near-equilibrium dynamics of hot gauge theory 
plasma there are several distinct ways to extract the transport coefficients \eqref{n4res}.
First \cite{u2}, the shear viscosity can be computed from the two-point 
correlation function of the stress-energy 
tensor at zero spatial momentum via the Kubo formula 
\begin{equation}
\eta=\lim_{\w\to 0} \frac{1}{2\w}\int dtd\bar{x}\ e^{i\w t}\langle\left[T_{xy}(x),\ T_{xy}(0)\right]\rangle\,.
\eqlabel{kubo}
\end{equation} 
Second \cite{pol1}, diffusive channel two-point retarded correlation 
function of the stress energy tensor, for example,
\begin{equation}
\begin{split}
G_{tx,tx}(\w,q)=-i \int dt d\bar{x} e^{-i\w t+i q z}\theta(t)   
\langle\left[T_{tx}(x),\ T_{tx}(0)\right]\rangle
\propto  \frac{1}{i\w-\cald q^2}\,,
\end{split}
\eqlabel{shearchan}
\end{equation} 
have a pole at 
\begin{equation}
\w=-i \cald q^2\,,
\eqlabel{diffpole}
\end{equation}
where the shear diffusion constant $\cald$ is 
\begin{equation}
\cald=\frac{\eta}{s T}\,.
\eqlabel{ddef}
\end{equation}
 Finally, all the transport coefficients 
\eqref{n4res} can be read off from the dispersion relation of a pole in the sound wave 
channel two-point retarded correlation function of the stress energy tensor, for example,
\begin{equation}
\begin{split}
G_{tt,tt}(\w,q)=-i \int dt d\bar{x} e^{-i\w t+i q z}\theta(t)   \langle\left[T_{tt}(x),\ T_{tt}(0)\right]\rangle\,,
\end{split}
\eqlabel{soundchan}
\end{equation}
as 
\begin{equation}
\w(q)=c_s q -i\ \frac{2q^2}{3 T }\ \frac{\eta}{s}\ \left(1+\frac{3\zeta}{4\eta}\right) \,.
\eqlabel{sounddisp}
\end{equation}
In \cite{sh1}, using the Kubo formula, the finite 't Hooft coupling correction to the ratio \eqref{etas} 
was computed. Such corrections were further extracted from the two-point retarded correlation functions 
\eqref{shearchan} and \eqref{soundchan} in \cite{sh2}. Much like as for the infinite 't Hooft  coupling,
all three approaches led to the same result:
\begin{equation}
\frac{\eta}{s}=\frac{1}{4\pi}\left(1+\frac{135}{8}\ \zeta(3)\ \l^{-3/2}+\cdots\right)\,.
\eqlabel{etacor}
\end{equation}

The computation leading to \eqref{etacor} were performed in the supergravity approximation to type IIb string 
theory including the leading $\a'$ correction
\begin{equation}
I=  \frac{1}{ 16\pi G_{10}} \int d^{10} x \sqrt {-g}
\ \bigg[ R - {\frac 12} (\partial \phi)^2 - \frac{1}{4 \cdot 5!}   (F_5)^2  +...+ 
\  \gamma \ e^{- {\frac 3 2} \phi}  W + ...\bigg]   \  ,
\eqlabel{aaa}
\end{equation}
where
$$ \ \ \ \ \ \   
  \gamma= { \frac 18} \zeta(3)(\alpha')^3 \ , 
$$ 
corresponding to 
\begin{equation}
\gamma= { \frac 18} \zeta(3)\ \l^{-3/2}
\eqlabel{ggauge}
\end{equation}
in terms of the SYM 't Hooft coupling $\l$,
and 
\begin{equation}
W =  C^{hmnk} C_{pmnq} C_{h}^{\ rsp} C^{q}_{\ rsk} 
 + {\frac 12}  C^{hkmn} C_{pqmn} C_h^{\ rsp} C^{q}_{\ rsk}\  . 
\label{rrrr}
\end{equation}
A somewhat uncontrollable approximation\footnote{This is related with difficulties of deriving $\a'$ corrected 
string theory effective action in the presence of nontrivial RR fluxes.} used  in \cite{sh1,sh2} (and earlier in \cite{gkt} ) 
was an  assumption that in a chosen scheme \eqref{aaa}, the 
self-dual $F_5$ form does not receive order $(\a')^3$
corrections. Though the agreement of the shear viscosity extracted from different boundary correlation functions at order 
$\calo(\ga)$ computationally appears to be rather dramatic\footnote{For example, using \eqref{soundchan}, \eqref{sounddisp} one has to 
take into account $\a'$ corrections to the black three brane temperature \cite{gkt}.}, a recent work \cite{met} suggests that such an agreement 
is automatically encoded in any gravitational model, perhaps similar to the way the thermodynamic relation between the 
energy ($\cale$), free energy ($\calf$) and the entropy densities $\calf=\cale- T s$ is encoded in the horizon geometries \cite{abk}.    

In \cite{j2} the ratio of the shear viscosity to the entropy density at infinite 't Hoof coupling was computed for the 
Bjorken flow of the $\caln=4$ SYM plasma, and the result was found to agree with \eqref{etas}. 
In this paper we compute leading order finite 't Hooft coupling correction to $\eta/s$ for the $\caln=4$ plasma 
in the Bjorken flow. We use string theory effective action \eqref{aaa}, which is expected to reproduce \eqref{etacor}
obtained from the equilibrium correlation function computations. Unfortunately, we report here a disagreement with 
\eqref{etacor}:
\begin{equation}
\frac{\eta}{s}\bigg|_{Bjorken\ flow}=\frac{1}{4\pi}\left(1+\frac{120}{8}\ \zeta(3)\ \l^{-3/2}+\cdots\right)\,.
\eqlabel{floweta}
\end{equation}

In principle, analysis are very straightforward:
\nxt we compute $\a'$ corrections to the gravitational background dual to boost invariant Bjorken expansion on the boundary;
\nxt as in \cite{j2}, we fix some of the parameters\footnote{Amusingly, not all parameters can 
be fixed from the nonsingularity condition \cite{j1}. Despite this, our answer for the 
viscosity ratio \eqref{floweta} is unambiguous.} of the $\a'$ corrected geometry by requiring nonsingularity of the string theory background
to order $\calo(\t^{-4/3})$ in the late proper time expansion;
\nxt we use $\a'$ corrected holographic renormalization (see \cite{bal}) to evaluate the energy density of the expanding $\caln=4$ plasma;
\nxt as in \cite{bbhj} we interpret the string theory computation in the framework of the Muller-Israel-Stewart theory  of dissipative relativistic 
hydrodynamics\footnote{In \cite{sl} a correction to Bjorken expansion within MIS theory was pointed out. Such a correction, while modify 
the interpretation of the plasma relaxation time, 
does not affect the determination of its shear viscosity compare to the analysis of \cite{bbhj}.}.

Given the computational complexity of the analysis and the disagreement we claim to exist between equilibrium correlation function computation and the 
Bjorken flow computation of the plasma shear viscosity at finite 't Hooft coupling, the remainder of the paper 
focuses on explaining the consistency checks we performed.    

\section{Computational framework}
The most difficult step in the analysis is to obtain $\a'$-corrected supergravity equations governing the SYM Bjorken flow at finite 
't Hooft coupling. As in \cite{bal}, we first derive effective action for the warp factors of the gravitation background, and then determine  
$\a'$-corrected equations of motion. As explained in \cite{bbhj}, choosing a Bjorken frame, fixes reparametrization of a radial and a proper time 
coordinates, which leads to two first order differential constraints (at each order in the late proper time expansion) on the warp factors. 
Not only these constraints are important consistency checks on the obtained system of equations, but they are also crucial to pick up correct 
solutions\footnote{These constraints fix some of the integration constants.}. If one naively obtains effective action for the warp factors,
the constraints are lost. We explain how to properly derive effective action, so that the reparametrization constraints are kept. 
As in the supergravity approximation, the $\a'$ corrected  constraints will provide vital consistency checks.     

\subsection{Supergravity approximation: effective action, constraints and the solution}
  
The supergravity background holographically dual to a Bjorken flow of the $\caln=4$ plasma takes form \cite{j1,bbhj}
\begin{equation}\eqlabel{10dimM}
 \begin{split}
  d\tilde{s}_{10}^{2}\:&=\:\tilde{g}_{MN}d\xi^{M}d\xi^{N}\:=\\
  &=\:e^{-2 \a(\t,z)}g_{\mu\nu}(x)dx^{\mu}dx^{\nu}+
      e^{6/5\a(\t,z)}\left(dS^{5}\right)^{2}\,,
 \end{split}
\end{equation}
 where $(dS^{5})^{2}$ is the line element for a $5$-dimensional sphere with unit
radius, and 
\begin{equation}\eqlabel{5dimM}
 \begin{split}
  ds^{2}\:&=\:g_{\mu\nu}dx^{\mu}dx^{\nu}\:=\:\\
          &=\frac{1}{z^{2}}
	    \left[
             -e^{2a(\tau,z)}d\tau^{2}+e^{2b(\tau,z)}\tau^{2}dy^{2}+
	      e^{2c(\tau,z)}dx_{\perp}^{2}
	    \right]+
	    \frac{dz^{2}}{z^{2}}\,,
 \end{split}
\end{equation}
where $dx_{\perp}^{2}\equiv dx_1^2+dx_2^2$.
The $5$-form $F_{5}$ takes form\footnote{We normalize the five-form flux so that the asymptotic AdS radius is one.}
\begin{equation}\eqlabel{5form}
F_{5}\:=\:\mathcal{F}_{5}+\star\mathcal{F}_{5}\,,\qquad
\mathcal{F}_{5}\:=\:-4\ \omega_{S^{5}}\,,
\end{equation}
where $\omega_{S^{5}}$ is the 5-sphere volume form.
Moreover,  the dilaton is $\phi=\phi(\t,z)$.

Notice that choosing the five-dimensional metric as in \eqref{5dimM} fixes the $(\t,z)$ repa-rametrization invariance 
so that 
\begin{equation}
g_{zz}=\frac{1}{z^2}\,,\qquad g_{\t z}=0\,.
\eqlabel{gauge}
\end{equation}
Substituting the ansatz \eqref{10dimM}-\eqref{5form} into type IIb supergravity action\footnote{As usual,
care should be taken with the self-dual five-form. The correct contribution to the effective action from the five-form is 
$-\frac{1}{4\cdot 5!}F_5^2=-8 e^{-6\a(\t,z)}$, see \cite{bbhj}.} ( the $\ga=0$ approximation of \eqref{aaa} )
one obtains naive effective action for the scalars $\{a,b,c,\a,\phi\}$:
\begin{equation}
S_{eff,naive}\biggl[a(\t,z),b(\t,z),c(\t,z),\a(\t,z),\phi(\t,z)\biggr]=I\bigg|_{\ga=0}\,.
\eqlabel{snaive}
\end{equation}
Clearly, from \eqref{snaive} one obtains only five second order PDE's, instead of seven, as in \cite{bbhj}. 
The reason that is happening is because the reparametrization constraints \eqref{gauge} are imposed 
directly on the action, instead on the equations of motion. 

In order the keep constraints within the effective action approach we deform the five dimensional metric \eqref{5dimM}
as follows
\begin{equation}\eqlabel{5dimMd}
 \begin{split}
  ds^{2}\:&=\:g_{\mu\nu}dx^{\mu}dx^{\nu}\:=\:\\
          &=\frac{1}{z^{2}}
	    \left[
             -e^{2a(\tau,z)}d\tau^{2}+e^{2b(\tau,z)}\tau^{2}dy^{2}+
	      e^{2c(\tau,z)}dx_{\perp}^{2}
	    \right]+h(\t,z)d\t dz+
	    \left(1+g(\t,z)\right)\frac{dz^{2}}{z^{2}}\,.
 \end{split}
\end{equation}
Constraints \eqref{gauge} now correspond to 
\begin{equation}
h(\t,z)\equiv0\,,\qquad g(\t,z)\equiv 0\,.
\eqlabel{gaugen}
\end{equation}
Next, we evaluate the supergravity action on \eqref{5dimMd}, \eqref{5form}
\begin{equation}
S_{eff}\biggl[a,b,c,\a,\phi\ ;\ h,g\biggr]=I\bigg|_{\ga=0}\,.
\eqlabel{seff}
\end{equation}
While it is important to maintain full nonlinear contributions for the ``physical`` scalars 
$\{a,b,c,\a,\phi\}$ in $S_{eff}$, it is sufficient to evaluate \eqref{seff} to linear order in $\{h,g\}$. 
The equations of motion derived from $S_{eff}$, followed by fixing the reparametrization invariance as in \eqref{gaugen}
\begin{equation}
\begin{split}
&0=\frac{\dd S_{eff}}{\dd a(\t,z)}\bigg|_{h(\t,z)=g(\t,z)=0}\,,\ 0=\frac{\dd S_{eff}}{\dd b(\t,z)}\bigg|_{h(\t,z)=g(\t,z)=0}\,,
\ 0=\frac{\dd S_{eff}}{\dd c(\t,z)}\bigg|_{h(\t,z)=g(\t,z)=0}\,,\\
&0=\frac{\dd S_{eff}}{\dd \a(\t,z)}\bigg|_{h(\t,z)=g(\t,z)=0}\,,\ 0=\frac{\dd S_{eff}}{\dd \phi(\t,z)}\bigg|_{h(\t,z)=g(\t,z)=0}\,,\\
&0=\frac{\dd S_{eff}}{\dd h(\t,z)}\bigg|_{h(\t,z)=g(\t,z)=0}\,,\qquad 0=\frac{\dd S_{eff}}{\dd g(\t,z)}\bigg|_{h(\t,z)=g(\t,z)=0}\,,
\end{split}
\eqlabel{eom}
\end{equation}
are precisely equivalent to the full set of equations derived in \cite{bbhj} (see eq.~(3.12)-(3.18)).

Equations \eqref{eom} are solved as a series expansion in the late proper time $\t\to \infty$, but exactly 
in the scaling variable \cite{j1}
\begin{equation}
v\equiv \frac{z}{\t^{1/3}}\,.
\eqlabel{vv}
\end{equation}   
We find it useful (especially when we generalize our effective action approach to the full action \eqref{aaa} )
to rewrite $S_{eff}$ as a functional of 
$$\{a(\t,v),\ b(\t,v),\ c(\t,v),\ \a(\t,v),\ \phi(\t,v),\ h(\t,v),\ g(\t,v)\}\,,$$ 
\ie, fields depending on the scaling variable \eqref{vv}. When we do this, it is important to note the modification of the 
effective action measure
$$dt dz\cdots\ \to\ t^{1/3}dt dv\cdots  \,.$$ 

Solving \eqref{eom} in the scaling variable $v$ 
and with the boundary conditions
\begin{equation}
\biggl\{
a(\t,v), b(\t,v),c(\t,v),\a(\t,v),\phi(\t,v)
\biggr\}\bigg|_{v\to 0}=0\,,
\eqlabel{bc}
\end{equation}
we find \cite{j1,j2}
\begin{equation}
\begin{split}
&a(\tau,v) = a_{0}(v) + \frac{1}{\tau^{2/3}} a_{1}(v) + \frac{1}{\tau^{4/3}} a_{2}(v) + \calo(\t^{-2})\,,\\
&b(\tau,v) = b_{0}(v) + \frac{1}{\tau^{2/3}} b_{1}(v) + \frac{1}{\tau^{4/3}} b_{2}(v) +\calo(\t^{-2})\,,\\
&c(\tau,v) = c_{0}(v) + \frac{1}{\tau^{2/3}} c_{1}(v) + \frac{1}{\tau^{4/3}} c_{2}(v) +\calo(\t^{-2})\,,\\
&\a(\t,v)=\calo(\t^{-2})\,,\qquad \phi(\t,v)=\calo(\t^{-2})\,,
\end{split}
\eqlabel{expansion1}
\end{equation} 
with 
\begin{equation}
\begin{split}
a_0=\frac 12 \ln\frac{\left(1-v^4/3\right)^2}{1+v^4/3}\,,\qquad b_0=c_0=\frac 12 \ln\left(1+v^4/3\right)\,,
\end{split}
\eqlabel{order0}
\end{equation}
\begin{equation}
\begin{split}
&a_1=\he\ \frac{\left(9-v^4\right)v^4}{9-v^8}\,,\qquad  c_1=-\he\ \frac{v^4}{3+v^4}-\frac{\he}{2}\ \ln\frac{3-v^4}{3+v^4}\,, \\
&b_1=-3\he\ \frac{v^4}{3+v^4}-2 c_1\,,
\end{split}
\eqlabel{order1}
\end{equation}
\begin{equation}
\begin{split}
a_2=&\frac{(9+5v^4)v^2}{12(9-v^8)} -C \frac{(9+v^4)v^4}{72(9-v^8)} +
\he^2 \frac{(-1053-171v^4+9v^8+7v^{12})v^4}{6(9-v^8)^2}\\
&+
\frac{1}{8\sqrt{3}} \ln \frac{\sqrt{3}-v^2}{\sqrt{3}+v^2}-
\frac{3}{4} \he^2 \ln \frac{3-v^4}{3+v^4}\,,  \\
c_2 =& -\frac{\pi^2}{288\sqrt{3}} +\frac{v^2(9+v^4)}{12(9-v^8)} +C
\frac{v^4}{72(3+v^4)} -\he^2 \frac{(-9+54v^4+7v^8)v^4}{6(3+v^4)(9-v^8)}\\
& +
\frac{1}{8\sqrt{3}} \ln \frac{\sqrt{3}-v^2}{\sqrt{3}+v^2}+\frac{1}{72} (C+66\he^2) \ln \frac{3-v^4}{3+v^4}\\
&+
 \frac{1}{24\sqrt{3}} \left( \ln \frac{\sqrt{3}-v^2}{\sqrt{3}+v^2}
\ln\frac{(\sqrt{3}-v^2)(\sqrt{3}+v^2)^3}{4(3+v^4)^2} -{\rm li}_2 \left(-
\frac{(\sqrt{3}-v^2)^2}{(\sqrt{3}+v^2)^2} \right)\right)\,,\\
b_2 =& -2 c_2+\frac{v^2}{4(3+v^4)} +C\frac{v^4}{24(3+v^4)} +\he^2
\frac{(39+7v^4)v^4}{2(3+v^4)^2} +
\frac{1}{8\sqrt{3}} \ln \frac{\sqrt{3}-v^2}{\sqrt{3}+v^2}\\
&+
\frac{3}{4} \he^2 \ln \frac{3-v^4}{3+v^4}\,,
\end{split}
\eqlabel{order2}
\end{equation}
for arbitrary parameters $\{\he,C\}$.

\subsection{Beyond the supergravity approximation}
To go beyond the supergravity approximation we deform proper late time expansion for the physical fields 
\eqref{expansion1} as
\begin{equation}
\begin{split}
&a(\tau,v) = \biggl(a_{0}(v)+\ga \ha_0(\t,v)\biggr) + \frac{1}{\tau^{2/3}}\biggl(a_{1}(v)+\ga \ha_1(\t,v)\biggr) + \frac{1}{\tau^{4/3}} 
\biggl(a_{2}(v)+\ga \ha_2(\t,v)\biggr)\,,\\
&b(\tau,v) = \biggl(b_{0}(v)+\ga \hb_0(\t,v)\biggr) + \frac{1}{\tau^{2/3}}\biggl(b_{1}(v)+\ga \hb_1(\t,v)\biggr) + \frac{1}{\tau^{4/3}} 
\biggl(b_{2}(v)+\ga \hb_2(\t,v)\biggr)\,,\\
&c(\tau,v) = \biggl(c_{0}(v)+\ga \hc_0(\t,v)\biggr) + \frac{1}{\tau^{2/3}}\biggl(c_{1}(v)+\ga \hc_1(\t,v)\biggr) + \frac{1}{\tau^{4/3}} 
\biggl(c_{2}(v)+\ga \hc_2(\t,v)\biggr)\,,\\
&\a(\tau,v) = \ga \hal_0(\t,v) + \frac{1}{\tau^{2/3}}\ \ga \hal_1(\t,v) + \frac{1}{\tau^{4/3}} 
\ga \hal_2(\t,v)\,,\\
&\phi(\tau,v) = \ga \hp_0(\t,v) + \frac{1}{\tau^{2/3}}\ \ga \hp_1(\t,v) + \frac{1}{\tau^{4/3}} 
\ga \hp_2(\t,v)\,,
\end{split}
\eqlabel{defform}
\end{equation}
where we assumed that as  $\l\to \infty$
\begin{equation} 
\begin{split}
&\ha_i(\l\t,v)\to \ha_i(\t,v)\,,\ \hb_i(\l\t,v)\to \hb_i(\t,v)\,,\ \hc_i(\l\t,v)\to \hc_i(\t,v)\,,\\
&\hal_i(\l\t,v)\to \hal_i(\t,v)\,,\ 
\hp_i(\l\t,v)\to \hp_i(\t,v)\,,\qquad i=0,1,2\,.
\end{split}
\eqlabel{scaling}
\end{equation}
Substituting \eqref{defform} into \eqref{aaa} we can obtain effective action for 
\begin{equation}
\{\ha_i,\hb_i,\hc_i,\hal_i,\hp_i\}\,.
\eqlabel{dfields}
\end{equation}
We need to expand the supergravity part of \eqref{aaa} to quadratic order in the fields \eqref{dfields}, while it is sufficient to 
evaluate the $\ga e^{-3/2 \phi} W$ term in \eqref{aaa} to linear order in \eqref{dfields}. As explained in the previous section, in order 
to obtain reparametrization constraints  we further need to 
evaluate the whole action \eqref{aaa} to linear order in $\{h,g\}$. 
Resulting effective action can be organized as follows
\begin{equation}
\begin{split}
&S_{eff}^\ga\biggl[\ha_i,\hb_i,\hc_i,\hal_i,\hp_i;h,g;\he,C\biggr]=\int 
d\t\int dv\ \call_{eff}^{\ga}\biggl[\ha_i,\hb_i,\hc_i,\hal_i,\hp_i;h,g;\he,C\biggr]\,,\\
&\call_{eff}^\ga=\biggl\{\ga^2 \call^{-1/3}_{[2]}+\ga \call^{-1/3}_{[1]}\biggr\} +\biggl\{\ga^2 \call^{-3/3}_{[2]}+\ga \call^{-3/3}_{[1]}\biggr\}
+\biggl\{\ga^2 \call^{-5/3}_{[2]}+\ga \call^{-5/3}_{[1]}\biggr\}\,,
\end{split}
\eqlabel{orders}
\end{equation}
where, given \eqref{scaling}, the superscript in $\call$ indicates the leading scaling behaviour as $\t\to \infty$, \ie, 
\begin{equation}
\call^q=\calo\left(\t^q\right)\,.
\eqlabel{calls}
\end{equation}
Notice that we suppressed explicit dependence on $\{\he,C\}$ in the decomposition of $\call_{eff}^\ga$.

In the rest of this subsection we present further decomposition of $\call^q$, and explain the 
structure of the consistency checks on the example of the equation of motion for $\ha_2$. 

In the following expressions all $\call$'s do not scale as $\t\to \infty$ --- the explicit scaling dependence is factored out.
A superscript indicates whether a term comes from the supergravity part of \eqref{aaa}, or from the $\calo(\ga)$-correction in \eqref{aaa}; 
the subscripts indicate order of fields on which a lagrangian term depends. For example, the subscript in 
$\call_{(2,1)}$
implies that $\call_{(2,1)}$ is strictly bilinear in order order-2 and order-1 fields:
$$\call_{(2,1)}=\call_{(2,1)}\biggl[\left(\ha_2,\hb_2,\hc_2,\hal_2,\hp_2\right),\left(\ha_1,\hb_1,\hc_1,\hal_1,\hp_1\right)\biggr]\,,$$
while the subscript in $\call_{(1,1)}$ implies that such a term is strictly quadratic in order-1 fields 
$$\call_{(1,1)}=\call_{(1,1)}\biggl[\left(\ha_1,\hb_1,\hc_1,\hal_1,\hp_1\right),\left(\ha_1,\hb_1,\hc_1,\hal_1,\hp_1\right)\biggr]\,.$$
A single subscript indicates the order of fields on which the corresponding lagrangian term depends linearly.
Finally, no subscript indicates that the lagrangian terms are independent of \eqref{dfields}, but are linear in 
either $h$ or $g$.
Notice that lagrangian terms  with a single subscript coming from the supergravity part of \eqref{aaa} are necessarily
linear\footnote{This is simply a check (which we verified to be true) that the supergravity background \eqref{order0}-\eqref{order1} solves supergravity 
equations of motion.} in either $h$ or $g$. The latter implies that reparametrization constraints at leading order in $\ga$ appear as $\calo(\ga)$, 
and thus in the lagrangian terms which are either bilinear or quadratic in the fields \eqref{dfields} we can set $h=g=0$.
For the same reason we can set $h=g=0$ in the lagrangian terms of order $\calo(\ga^2)$ coming  from the $\calo(\ga)$-correction in \eqref{aaa}.

\subsubsection{Order-0}
We find 
\begin{equation}
\begin{split}
&\call_{[2]}^{-1/3}=\frac{1}{\t^{1/3}}\ \biggl(\ ^{[-1/3]}\call_{(0,0)}^{SUGRA}+\ ^{[-1/3]}\call_{(0)}^{W}\biggr)\,,\\
&\call_{[1]}^{-1/3}=\frac{1}{\t^{1/3}}\ \biggl(\ ^{[-1/3]}\call_{(0)}^{SUGRA}+\ ^{[-1/3]}\call^{W}\biggr)\,.
\end{split}
\eqlabel{02}
\end{equation}

\subsubsection{Order-1}
We find 
\begin{equation}
\begin{split}
\call_{[2]}^{-3/3}=&\frac{1}{\t}\ \biggl(\ ^{[-3/3]}\call_{(0,1)}^{SUGRA}+\ ^{[-3/3]}\call_{(1)}^{W}\biggr)\\
&+\frac{1}{\t^{5/3}}\ \biggl(\ ^{[-5/3]}\call_{(0,1)}^{SUGRA}+\ ^{[-5/3]}\call_{(1,1)}^{SUGRA}+\ ^{[-5/3]}\call_{(1)}^{W}\biggr)\,,
\end{split}
\eqlabel{12}
\end{equation}
\begin{equation}
\begin{split}
\call_{[1]}^{-3/3}=\frac{1}{\t}\ \biggl(\ ^{[-3/3]}\call_{(0)}^{SUGRA}+\ ^{[-3/3]}\call_{(1)}^{SUGRA}+\ ^{[-3/3]}\call^{W}\biggr)\,.
\end{split}
\eqlabel{11}
\end{equation}

\subsubsection{Order-2}
We find 
\begin{equation}
\begin{split}
\call_{[2]}^{-5/3}=&\frac{1}{\t^{5/3}}\ \biggl(\ ^{[-5/3]}\call_{(0,2)}^{SUGRA}+\ ^{[-5/3]}\call_{(2)}^{W}\biggr)\\
&+\frac{1}{\t^{7/3}}\ \biggl(\ ^{[-7/3]}\call_{(0,2)}^{SUGRA}+\ ^{[-7/3]}\call_{(1,2)}^{SUGRA}+\ ^{[-7/3]}\call_{(2)}^{W}\biggr)\\
&+\frac{1}{\t^{9/3}}\ \biggl(\ ^{[-9/3]}\call_{(0,2)}^{SUGRA}+\ ^{[-9/3]}\call_{(1,2)}^{SUGRA}+\ ^{[-9/3]}\call_{(2,2)}^{SUGRA}+\ ^{[-9/3]}\call_{(2)}^{W}\biggr)\,,
\end{split}
\eqlabel{22}
\end{equation}
\begin{equation}
\begin{split}
\call_{[1]}^{-5/3}=\frac{1}{\t^{5/3}}\ \biggl(\ ^{[-5/3]}\call_{(0)}^{SUGRA}+\ ^{[-5/3]}\call_{(1)}^{SUGRA}+\ ^{[-5/3]}\call_{(2)}^{SUGRA}
+\ ^{[-5/3]}\call^{W}\biggr)\,.
\end{split}
\eqlabel{21}
\end{equation}

\subsubsection{Equations of motion for $\ha_2$ and consistency check on lower order solution}
Consider a variation
\begin{equation}
EOM[\ha_2]\equiv \biggl\{\frac{\dd}{\dd \ha_2(\t,v)}\ \int dt\int dv\ \call_{eff}^\ga\biggr\}\bigg|_{h=g=0}\,.
\eqlabel{vara2}
\end{equation}
Given  \eqref{orders}, \eqref{22}, we have\footnote{ The term $\ ^{[-5/3]}\call_{(2)}^{SUGRA}$ does not contribute as it can only be bilinear in 
both $\ha_2$ and either $h$ or $g$.} 
\begin{equation}
\begin{split}
EOM[\ha_2]=&\frac{1}{\t^{5/3}}\biggl(\ ^{[-5/3]}\cali_{(0)}+\ ^{[-5/3]}\cali\biggr)\\
&+\frac{1}{\t^{7/3}}\ \biggl(\ ^{[-7/3]} \cali_{(0)}+\ ^{[-7/3]} \cali_{(1)}+\ ^{[-7/3]} \cali\biggr)\\
&+\frac{1}{\t^{9/3}}\ \biggl(\ ^{[-9/3]} \cali_{(0)}+\ ^{[-9/3]} \cali_{(1)}+\ ^{[-9/3]} \cali_{(2)}+\ ^{[-9/3]} \cali\biggr)\,,
\end{split}
\eqlabel{valr}
\end{equation}
with obvious notations. For example:
\begin{equation}
\frac{1}{\t^{7/3}}\ ^{[-7/3]} \cali_{(0)}=\frac{\dd}{\dd \ha_2(\t,v)}\ \int dt\int dv\  \biggl\{ 
\frac{1}{\t^{7/3}}\ \ ^{[-7/3]}\call_{(0,2)}^{SUGRA}\biggr\}\,.
\end{equation}
Thus, a single variation \eqref{vara2} not only will produce the equation of motion for $\ha_2$, namely
\begin{equation}
0=\ ^{[-9/3]} \cali_{(0)}+\ ^{[-9/3]} \cali_{(1)}+\ ^{[-9/3]} \cali_{(2)}+\ ^{[-9/3]} \cali\,,
\eqlabel{eomha2t}
\end{equation}
but also would produce a consistency tests on the order-0 and order-1 solutions:
\begin{equation}
\begin{split}
0=&\ ^{[-5/3]}\cali_{(0)}+\ ^{[-5/3]}\cali\,,\\
0=&\ ^{[-7/3]} \cali_{(0)}+\ ^{[-7/3]} \cali_{(1)}+\ ^{[-7/3]} \cali\,,
\end{split}
\eqlabel{2tests}
\end{equation}
correspondingly.
It is easy to understand the origin of consistency constraints \eqref{2tests}: a field deformation at order-2 can be considered as 
a variation of a field deformation at order-0 or order-1; thus, once equations of motion at lower orders are solved, such variations 
must vanish.

Altogether, there are 10 consistency constraints when deriving equations for order-2 field deformations, and 5 consistency constraints 
when deriving equations for order-1 field deformations. We explicitly verified that all such constraints are satisfied.  

\section{Equations of motion and solutions for \eqref{dfields}}

We obtain equations of motion (including the constraints) at each order $i=0,1,2$ for \eqref{dfields} as explained in the previous section.
It is important that in deriving these equations of motion a field is assumed to depend both on $\{\t,v\}$. Once the equations of motion
are computed, we can assume 
$$
\{\ha_i,\hb_i,\hc_i,\hal_i,\hp_i\}(\t,v)\equiv \{\ha_i,\hb_i,\hc_i,\hal_i,\hp_i\}(v)\,.
$$
All the equations must be solved with the boundary conditions
\begin{equation}
\biggl\{\ha_i(v),\hb_i(v),\hc_i(v),\hal_i(v),\hp_i(v)\biggr\}\bigg|_{v\to 0}=0\,.
\eqlabel{bbcc}
\end{equation}

\subsection{Order-0}
To leading order we find the following system of equations
\begin{equation}
\begin{split}
0=&\hc_0''+\frac 12 \hb_0''+\frac{5 v^4-9}{(3+v^4) v} \hc_0'+ \frac{5 v^4-9}{2(3+v^4) v} \hb_0'
+\frac{933120 (4 v^8-33 v^4+36) v^{10}}{(3+v^4)^8}\,,
\end{split}
\eqlabel{eom01}
\end{equation}
\begin{equation}
\begin{split}
0=&\hc_0''+\frac 12 \ha_0''+\frac{5 v^4+9}{2(v^4-3) v} \ha_0'+\frac{5 v^8+27}{v (v^8-9)} \hc_0-\frac{311040 v^{10} (4 v^8-45 v^4+36)}{(3+v^4)^8}\,,
\end{split}
\eqlabel{eom02}
\end{equation}
\begin{equation}
\begin{split}
0=&\hc_0''+\ha_0''+\hb_0''+\frac{5 v^4+9}{(v^4-3) v} \ha_0'+\frac{5 v^8+27}{v (v^8-9)} \hb_0'
+\frac{5 v^8+27}{v (v^8-9)} \hc_0'\\
&-\frac{622080 v^{10} (4 v^8-45 v^4+36)}{(3+v^4)^8}\,,
\end{split}
\eqlabel{eom03}
\end{equation}
\begin{equation}
\begin{split}
0=&\hc_0''+\frac 12 \hb_0''+\frac{3 (v^8-10 v^4-3)}{v (v^8-9)} \hc_0'-\frac{3 (5 v^4-3)}{v (v^8-9)} \hb_0'-\frac{3(v^4-3)}{2(3+v^4) v} \ha_0'\\
&
+\frac{3732480 v^{10} (v^4-3)^2}{(3+v^4)^8}\,,
\end{split}
\eqlabel{eom04}
\end{equation}
\begin{equation}
\begin{split}
0=&\hc_0'+\frac 12 \hb_0'+\frac{(v^4-3)^2}{2(v^4+2 v^2+3) (v^4-2 v^2+3)} \ha_0'\\
&-\frac{2799360 (v^4-3) v^{15}}{(v^4-2 v^2+3) (v^4+2 v^2+3) (3+v^4)^7}\,,
\end{split}
\eqlabel{eom05}
\end{equation}
\begin{equation}
\begin{split}
0=&\hal_0''+\frac{5 v^8+27}{v (v^8-9)} \ha_0'-\frac{32}{v^2} \ha_0+\frac{2332800 v^{14}}{(3+v^4)^8}\,,
\end{split}
\eqlabel{eom06}
\end{equation}
\begin{equation}
\begin{split}
0=&\hp_0''+\frac{5 v^8+27}{v (v^8-9)}\hp_0'-\frac{5598720 v^{14}}{(3+v^4)^8}\,.
\end{split}
\eqlabel{eom07}
\end{equation}
Although the system \eqref{eom01}-\eqref{eom07} is overdetermined (we have 7 ODE's for the 5 functions), 
it is straightforward to verify that it is consistent.

Most general solution to \eqref{eom01}-\eqref{eom07}, subject to the boundary conditions \eqref{bbcc}, 
is parametrized by three arbitrary integration constants\footnote{Altogether we expect 10 integration constants; setting nonnormalizable components 
of the fields to zero 
fixes 5 integration constants; the two constraints fix another 2 integration constants.} $\{\dd_1,\b_1,\b_2\}$: 
\begin{equation}
\begin{split}
\ha_0=&\frac{72576} {( 3+{v}^{4} ) ^{3}}-\frac{585144} {( 3+{v}^{4}
 ) ^{4}}-\frac{216} { 3+{v}^{4}}+\frac{1714608} {( 3
+{v}^{4} ) ^{5}}-\frac{864} {( 3+{v}^{4} ) ^{2}}-\frac{1714608}
{ ( 3+{v}^{4} ) ^{6}}\\
&+ \left(  \frac{1}{12+4{v}^{4}  
}+\frac{1}{ 2{v}^{4}-6} \right)  ( 288+\dd_1 ) +24+\frac{1}{12}\dd_1\,,
\end{split}
\eqlabel{da0}
\end{equation}
\begin{equation}
\begin{split}
\hb_0=\frac{223560} {( 3+{v}^{4} ) ^{4}}-\frac{25920} {( 3+{v}^{4}
 ) ^{3}}+\frac{664848} {( 3+{v}^{4} ) ^{6}}-\frac{664848}
{ ( 3+{v}^{4} ) ^{5}}-{\frac {{\dd_1}}{4(3+{v}^{4})}}
+24+\frac{1}{12}\dd_1\,,
\end{split}
\eqlabel{db0}
\end{equation}
\begin{equation}
\hc_0=\hb_0\,,
\eqlabel{dc0}
\end{equation}
\begin{equation}
\begin{split}
\hal_0=&\beta_1 \left(\left(\frac{v^4}{864}+\frac{1}{96 v^4}\right) \ln\frac{3-v^4}{3+v^4}+\frac{1}{144}\right)+\frac{25}{16} 
\left(v^4+\frac{9}{v^4}\right) \ln\frac{3-v^4}{3+v^4}+\frac{75}{8}\\
&+\frac{225}{2(3+v^4)^2}+\frac{675}{(3+v^4)^3}-\frac{22275}{2(3+v^4)^4}+\frac{36450}{(3+v^4)^5}-\frac{36450}{(3+v^4)^6}\,,
\end{split}
\eqlabel{dal0}
\end{equation}
\begin{equation}
\begin{split}
\hp_0=&\left(\frac{\beta_2}{24}-\frac{45}{4}\right) \ln\frac{3-v^4}{3+v^4}
-\frac{135}{2(3+v^4)}-\frac{405}{2(3+v^4)^2}-\frac{810}{(3+v^4)^3}+\frac{25515}{(3+v^4)^4}
\\
&-\frac{87480}{(3+v^4)^5}+\frac{87480}{(3+v^4)^6}\,.
\end{split}
\eqlabel{dph0}
\end{equation}

For a generic choice of $\b_1$ the curvature invariants of the ten-dimensional geometry \eqref{10dimM} will be singular
as $v\to 3^{1/4}_-$. For example, 
\begin{equation}
\calr_{\mu\nu\rho\l}\calr^{\mu\nu\r\l}=-\frac{14}{3}\ga\ (1350+\b_1)\ln(3-v^4)+{\rm finite}\,.
\eqlabel{riem20}
\end{equation}
However, once we chose 
\begin{equation}
\b_1=-1350\,,
\eqlabel{b1c}
\end{equation}
the curvature invariant \eqref{riem20} is finite. We verified that for \eqref{b1c} all other curvature invariants 
(including those discussed in \cite{bbhj}) are finite.

Notice that unlike studies of the string theory dual of the Bjorken flow in the supergravity approximation 
\cite{j1,j2,j3,bbhj}, here the background geometry is not completely fixed from the nonsingularity 
condition --- two arbitrary integration constants $\{\dd_1,\b_2\}$ remain. Though $\b_2$ does not affect 
the holographic stress energy tensor of the background geometry, the latter does depend on $\dd_1$. 
Additional arbitrariness will appear at higher orders, however, the string theory computation when properly matched to the 
near-equilibrium SYM dynamics provides unambiguous predictions for plasma transport properties. We further 
comment on the physical origin of the arbitrariness in section 5.

\subsection{Order-1}
Using results of the previous subsection, including \eqref{b1c}, we find the following set of equations for the 
next-to-leading order in the late proper time  expansion at order $\calo(\ga)$
\begin{equation}
\begin{split}
0=&\hc_1''+\frac 12 \hb_1''+\frac{5 v^4-9}{(3+v^4) v} \hc_1'+\frac{5 v^4-9}{2(3+v^4) v} \hb_1'+\cals_{(1,1)}\,,
\end{split}
\eqlabel{eom11}
\end{equation}
\begin{equation}
\begin{split}
0=&\hc_1''+\frac 12 \ha_1''+ \frac{5 v^4+9}{2(v^4-3) v} \ha_1'+\frac{5 v^8+27}{v (v^8-9)} \hc_1'+\cals_{(1,2)}\,,
\end{split}
\eqlabel{eom12}
\end{equation}
\begin{equation}
\begin{split}
0=&\hc_1''+\ha_1''+\hb_1''+\frac{5 v^4+9}{(v^4-3) v} \ha_1'+\frac{5 v^8+27}{v (v^8-9)} \hb_1'
+\frac{5 v^8+27}{v (v^8-9)} \hc_1'+\cals_{(1,3)}\,,
\end{split}
\eqlabel{eom13}
\end{equation}
\begin{equation}
\begin{split}
0=&\hc_1''+\frac 12 \hb_1''+\frac{(v^4-15) v^3}{v^8-9}\hb_1'+\frac{5 v^8-30 v^4-27}{v (v^8-9)} \hc_1'
-\frac{ 3(v^4-3)}{2(3+v^4) v} \ha_1'-\frac{24 v^2}{v^8-9} \hb_1\\
&-\frac{48 v^2}{v^8-9} \hc_1+\cals_{(1,4)}\,,
\end{split}
\eqlabel{eom14}
\end{equation}
\begin{equation}
\begin{split}
0=&\hc_1'+ \frac{(v^4-3)^2}{2(v^4-2 v^2+3) (v^4+2 v^2+3)} \ha_1'+\frac 12 \hb_1'+\cals_{(1,5)}\,,
\end{split}
\eqlabel{eom15}
\end{equation}
\begin{equation}
\begin{split}
0=&\hal_1''+\frac{5 v^8+27}{v (v^8-9)} \hal_1'-\frac{32}{v^2} \hal_1+\cals_{(1,6)}\,,
\end{split}
\eqlabel{eom16}
\end{equation}
\begin{equation}
\begin{split}
0=&\hp_1''+\frac{5 v^8+27}{v (v^8-9)} \hp_1'+\cals_{(1,7)}\,,
\end{split}
\eqlabel{eom17}
\end{equation}
where the source terms $\{\cals_{(1,1)},\cdots,\cals_{(1,7)}\}$ are given in Appendix A. 
As before, while the system \eqref{eom11}-\eqref{eom17} is overdetermined, we explicitly verified that it is 
consistent. 

Most general solution to \eqref{eom11}-\eqref{eom17}, subject to the boundary conditions \eqref{bbcc}, is parametrized by three new arbitrary 
constants $\{\dd_2,\b_3,\b_4\}$: 
\begin{equation}
\begin{split}
\ha_1=&-\frac{72013536 \he}{(3+v^4)^6}+\frac{61725888 \he}{(3+v^4)^7}+\frac{31189536 \he}{(3+v^4)^5}+\frac{3 \he (288+\dd_1)}{(v^4-3)^2}
-\frac{5987520 \he}{(3+v^4)^4}\\
&- \frac{3\he (1728+\dd_1)}{2(3+v^4)^2}+ \frac{576 \he+\frac 23 \dd_2}{2(v^4-3)}+ \frac{\frac 13 \dd_2-576 \he}{2(3+v^4)}
+\frac{445824 \he}{(3+v^4)^3}-\frac 16 \he \dd_1+\frac{\dd_2}{18} \,,
\end{split}
\eqlabel{da1}
\end{equation}
\begin{equation}
\begin{split}
\hb_1=&-\left(120 \he-\frac 16 \he \dd_1+\frac{\dd_2}{18}\right) \ln\frac{3-v^4}{3+v^4}-\frac{864 \he-\frac 12 \he \dd_1+\frac{\dd_2}{6}}{3+v^4}
+\frac{144 \he+\frac 12 \he \dd_1}{v^4-3}-\frac{3024 \he}{(3+v^4)^2}\\
&+\frac{3 \he \dd_1}{2(3+v^4)^2}-\frac{11687328 \he}{(3+v^4)^5}-\frac{107136 \he}{(3+v^4)^3}-\frac{23934528 \he}{(3+v^4)^7}+\frac{2091744 \he}
{(3+v^4)^4}+\frac{27597024 \he}{(3+v^4)^6}\\
&-\frac 16 \he \dd_1+\frac{\dd_2}{18}\,,
\end{split}
\eqlabel{db1}
\end{equation}
\begin{equation}
\begin{split}
\hc_1=&\left(60 \he+\frac{\dd_2}{36}-\frac{1}{12} \he \dd_1\right) \ln\frac{3-v^4}{3+v^4}-\frac{72 \he+\frac 14 \he \dd_1}{(v^4-3)}
+\frac{432 \he-\frac{\dd_2}{6}-\frac 14 \he \dd_1}{(v^4+3)}-\frac 16 \he \dd_1\\
&-\frac{179712 \he}{(3+v^4)^3}-\frac{23934528 \he}{(3+v^4)^7}
+\frac{2336688 \he}{(3+v^4)^4}+\frac{28086912 \he}{(3+v^4)^6}+\frac{\left(1512+\frac 32 \dd_1\right) \he}{(3+v^4)^2}\\
&-\frac{12177216 \he}{(3+v^4)^5}
+\frac{\dd_2}{18}\,,
\end{split}
\eqlabel{dc1}
\end{equation}
\begin{equation}
\begin{split}
\hal_1=&-\b_3 \left(\left(\frac{v^4}{864}+\frac{1}{96 v^4}\right) \ln\frac{3-v^4}{3+v^4}+\frac{1}{144}\right)+
\frac{25\he}{4}  \left(v^4+\frac{9}{v^4}\right) \ln\frac{3-v^4}{3+v^4}+\frac{75\he}{2}\\
&+\frac{450 \he v^8 (v^4-3) (v^8+24 v^4+9)}{(3+v^4)^7}\,,
\end{split}
\eqlabel{daalpha1}
\end{equation}
\begin{equation}
\begin{split}
\hp_1=&\left(\frac{\b_4}{24}-\frac{\b_2\he}{12} \right) \ln\frac{3-v^4}{3+v^4}-\frac  {\he \b_2}{4(3+v^4)}-\frac{ \he \b_2}{4(v^4-3)}
-\frac{1487160 \he}{(3+v^4)^5}-\frac{3149280 \he}{(3+v^4)^7}\\
&-\frac{405 \he}{(3+v^4)^2}-\frac{135 \he}{2(3+v^4)}+\frac{135 \he}{2(v^4-3)}
-\frac{2430 \he}{(3+v^4)^3}+\frac{218700 \he}{(3+v^4)^4}+\frac{3674160 \he}{(3+v^4)^6}\,.
\end{split}
\eqlabel{dph1}
\end{equation}

For a generic choice of $\b_3$ the curvature invariants of the ten-dimensional geometry \eqref{10dimM} will be singular
as $v\to 3^{1/4}_-$. For example, 
\begin{equation}
\calr_{\mu\nu\rho\l}\calr^{\mu\nu\r\l}=\frac{1}{\t^{2/3}}\ \frac{14}{3}\ga\left(\beta_3-5400\he\right)\ln(3-v^4)+{\rm finite}\,.
\eqlabel{riem21}
\end{equation}
However, once we chose 
\begin{equation}
\b_3=5400\he\,.
\eqlabel{b3c}
\end{equation}
the curvature invariant \eqref{riem21} is finite. We verified that for \eqref{b3c} all other curvature invariants 
(including those discussed in \cite{bbhj}) are finite.

At this stage we have four arbitrary integration constants: 
$\{\dd_1,\b_2,\dd_3,\b_4\}$.

\subsection{Order-2}
Using results of the previous two subsections, including \eqref{b1c} and \eqref{b3c}, we find the following set of equations for the next-to-next-to-leading 
order in the late proper time expansion at order $\calo(\ga)$
\begin{equation}
\begin{split}
0=&\hc_2''+\frac 12 \hb_2''+\frac{5 v^4-9}{(3+v^4) v} \hc_2'+\frac {5 v^4-9}{2(3+v^4) v} \hb_2'+\cals_{(2,1)}\,,
\end{split}
\eqlabel{eom21}
\end{equation} 
\begin{equation}
\begin{split}
0=&\hc_2''+\frac 12 \ha_2''+\frac{5 v^8+27}{v  (v^8-9)} \hc_2'+ \frac{9+5 v^4}{2v (v^4-3)} \ha_2'+\cals_{(2,2)}\,,
\end{split}
\eqlabel{eom22}
\end{equation} 
\begin{equation}
\begin{split}
0=&\hc_2''+\ha_2''+\hb_2''+\frac{9+5 v^4}{v (v^4-3)} \ha_2'+\frac{5 v^8+27}{v  (v^8-9)} \hb_2'+\frac{5 v^8+27}{v (v^8-9)} \hc_2'+\cals_{(2,3)}\,,
\end{split}
\eqlabel{eom23}
\end{equation} 
\begin{equation}
\begin{split}
0=&\hc_2''+\frac 12 \hb_2''- \frac{3(v^4-3)}{2(3+v^4) v} \ha_2'+\frac{2 v^8-15 v^4-9}{v  (v^8-9)} \hb_2'+\frac{7 v^8-30 v^4-45}{v (v^8-9)} \hc_2'
-\frac{48 v^2}{v^8-9} \hb_2\\
&-\frac{96 v^2}{v^8-9} \hc_2+\cals_{(2,4)}\,,
\end{split}
\eqlabel{eom24}
\end{equation} 
\begin{equation}
\begin{split}
0=&\hc_2'+\frac 12 \hb_2'+\frac{(v^4-3)^2}{2(v^4-2 v^2+3) (v^4+2 v^2+3)} \ha_2'+\cals_{(2,5)}\,,
\end{split}
\eqlabel{eom25}
\end{equation}
\begin{equation}
\begin{split}
0=&\hal_2''+\frac{5 v^8+27}{v  (v^8-9)} \hal_2'-\frac{32}{v^2} \hal_2+\cals_{(2,6)}\,,
\end{split}
\eqlabel{eom26}
\end{equation} 
\begin{equation}
\begin{split}
0=&\hp_2''+\frac{5 v^8+27}{v (v^8-9)} \hp_2'+\cals_{(2,7)}\,,
\end{split}
\eqlabel{eom27}
\end{equation} 
where the source terms $\{\cals_{(2,1)}\cdots,\cals_{(2,7)}\}$ are given in Appendix B.
As in previous subsections, while the system \eqref{eom21}-\eqref{eom27} is overdetermined, we explicitly verified that 
it is consistent. 

Solving \eqref{eom21}-\eqref{eom27} is quite complicated. Fortunately, we do not need a complete solution. 
Our ultimate goal is to determine $\he$ from the nonsingularity of the ten dimensional metric curvature invariants
to order $\calo(\ga)$ and to order $\calo(\t^{-4/3})$ in the late proper time expansion. Thus we evaluate 
metric invariants first, find  what field combinations affect the singularity as $v\to 3^{1/4}_-$, and then solve just for 
those combinations of field.
  
Up to now, all our equations and solutions are exact in $\he$. We assume now that 
\begin{equation}
\he=\he_0+\ga \he_1+\calo(\ga^2)\,,
\eqlabel{etaex}
\end{equation}
and evaluate background curvature invariants to order $\calo(\ga)$ near 
\begin{equation}
x\equiv 3^{1/4}-v\,.
\eqlabel{ydef}
\end{equation} 
We use explicit solutions \eqref{order2}, \eqref{da0}-\eqref{dph0} and \eqref{da1}-\eqref{dph1}
to evaluate curvature invariants. At this stage we use equations of motion \eqref{eom21}-\eqref{eom27} 
to eliminate the derivatives (if possible) of $\{\ha_2,\hb_2,\hc_2,\hal_2\}$ from the curvature invariants.

\subsubsection{$\calr$}
For  Ricci scalar we find (there is no dependence on order-2 fields here)
\begin{equation}
\begin{split}
\calr=-\frac 32\ \ga\ \frac{18\he_0^2-\sqrt{3}}{\t^{4/3}}\ \biggl(\frac{1}{x^4}-\frac{2}{3^{1/4}}\ \frac{1}{x^3}-\frac{13\sqrt{3}}{6}\ \frac{1}{x^2}
+\frac{3^{1/4}\ 5 }{2}\ \frac 1x\biggr)+{\rm finite}\,,\qquad x\to 0_+\,.
\end{split}
\eqlabel{rsc2}
\end{equation}
From \eqref{rsc2}  we  find that  Ricci scalar of the string theory geometry is nonsingular as $x\to 0_+$ when 
\begin{equation}
\he_0=\frac{1}{2^{1/2}3^{3/4}}\,,
\eqlabel{eta0}
\end{equation}
which is precisely the condition found from the nonsingularity of Riemann tensor squared \cite{j2}, as well as higher curvature invariants \cite{bbhj},
in the supergravity approximation to the string theory dual of the $\caln=4$ SYM Bjorken flow. The difference here (compare to \cite{j2,bbhj}) 
is that $\he_0$ is already fixed by requiring the nonsingularity of the Ricci scalar\footnote{Notice that Ricci scalar vanishes in the 
supergravity approximation \cite{bbhj}.}.

\subsubsection{$\calr_{\mu\nu\r\l}\calr^{\mu\nu\r\l}$}
A bit more work is necessary to determine the nonsingularity condition of Riemann tensor squared at order $\calo(\ga)$.
Generalizing the notation of \cite{bbhj}
\begin{equation}
\begin{split}
\cali^{[2]}&\equiv \calr_{\mu\nu\r\l}\calr^{\mu\nu\r\l}\\
&=\biggl(\cali^{[2]SUGRA}_0(v)+\ga\ \cali^{[2]W}_0(v)\biggr)
+\frac{1}{\t^{2/3}} \biggl(\cali^{[2]SUGRA}_1(v)+\ga\ \cali^{[2]W}_1(v)\biggr)
\\
&+\frac{1}{\t^{4/3}} \biggl(\cali^{[2]SUGRA}_2(v)+\ga\ \cali^{[2]W}_2(v)\biggr)+\calo(\t^{-2})+\calo(\ga^2)\,.
\end{split}
\eqlabel{i2def}
\end{equation}
With \eqref{eta0} and the lower order solutions, all terms in the decomposition of $\cali^{[2]}$ in \eqref{i2def} except for 
$\cali^{[2]W}_2(v)$ are finite as $v\to 3^{1/4}_-$. We find
\begin{equation}
\begin{split}
\cali^{[2]W}_2=&-\frac{2304 v^5 (v^4-3)}{(3+v^4)^3} f_2'(v)-\frac{192 (5 v^{16}+60 v^{12}+54 v^8+540 v^4+405)}{(3+v^4)^4} \hal_2(v)\\
&+\dd\cali_2(v)\,,
\end{split}
\eqlabel{riemann22}
\end{equation}
where 
\begin{equation}
f_2(v)=\hc_2(v)+\frac 12\ \hb_2(v)\,,
\eqlabel{deff2}
\end{equation} 
and $\dd\cali_2(v)$ does not depend on order-2 fields, and can be evaluated explicitly
\begin{equation}
\begin{split}
\dd\cali_2=&\biggl(-399\sqrt {3}+\frac{5\sqrt {3}}{6}{\dd_1}-\frac{2^{3/2}}{3^{3/4}}{
\dd_2}+ {2^{5/2}}{3^{5/4}}{\he_1}\biggr)\times\biggl(\frac{1}{x^4}-\frac{2}{3^{1/4}}\ \frac{1}{{x}^{3}}-\frac{\sqrt{3}}{6}
\ \frac{1}{x^2}\\
&+\frac {3^{1/4}}{2}{\frac {\left( 72\sqrt {2}\ {3}^{3/4}\ {
\he_1}-4\sqrt {2}\ {3}^{3/4}\ {\dd_2}+39\ {\dd_1}-270 \right) 
}{ \left( -7182+15\ {\dd_1}-4\sqrt {2}\ {3}^{3/4}\ {\dd_2}+72
\sqrt {2}\ {3}^{3/4}\ {\he_1} \right) }}\ \frac 1x
\biggr)+{\rm finite}\,,
\end{split}
\eqlabel{singddi2}
\end{equation}
as $x\to 0_+$.

From \eqref{eom21} one can obtain a decoupled equation for $f_2$ as defined in \eqref{deff2}.
The resulting equation is possible to explicitly solve for $f'_2$:
\begin{equation}
\begin{split}
f_2'(v)=&-\frac{24126004224 \he^2 v^3}{(v^4+3)^9}+\frac{13961808  v^3}{(v^4+3)^8}(2 v^2+1932 \he^2-C)
-\frac{34992  v^3}{(v^4+3)^7}(665 v^2\\
&+320520 \he^2-399 C)
+\frac{243  v^3}{(v^4+3)^6}(28221 v^2+8578560 \he^2-20600 C)
\\
&-\frac{81  v^3}{2(v^4+3)^5}(20823 v^2+3918720 \he^2-18560 C)+\frac{27  v^3}{8(v^4+3)^4}(7561 v^2+48 \dd_1 \he^2\\
&+921600 \he^2-11520 C)
-\frac{ v^3}{16(v^4+3)^3} (-26307 v^2+8 v^2 \dd_1-4 \dd_1 C+407808 \he^2\\
&+192 \he \dd_2-48 \dd_1 \he^2))+\frac{v^3}{(v^4+3)^2} \biggl(
\frac{35831808 \he^2-1990656 v^2}{(v^4-3)^5}\\
&+\frac{-1327104 v^2+29859840 \he^2}{(v^4-3)^4}+\frac{(8709120 \he^2-290304 v^2)}{(v^4-3)^3}
\\
&+\frac{1}{(v^4-3)^2}\left(-23472 v^2+1029024 \he^2-\frac 32 v^2 \dd_1-27 \dd_1 \he^2\right)
\\
&+\frac{1}{v^4-3}\left(9 \dd_1 \he^2-\frac 12 v^2 \dd_1+25488 \he^2-6 \he \dd_2+\frac{333}{2} v^2\right)\\
&+\frac{10425\sqrt{3}}{16}\ \arctan\ \frac{v^2}{\sqrt{3}}
+\dd_3\biggr)\,,
\end{split}
\eqlabel{df2}
\end{equation}
where $\dd_3$ is a new arbitrary integration constant.  
The solution \eqref{df2} is exact in $\he$.

Unfortunately, we can not present an explicit solution for $\hal_2(v)$. We can prove though that $\hal_2(v)$
can be chosen to be  finite as $v\to 3^{1/4}$, while having a vanishing nonnormalizable mode as $v\to 0_+$.
Indeed, the most general inhomogeneous solution  of \eqref{eom27} as $x\to 0_+$ takes form
\begin{equation}
\ha_2(x)=\calc_0+\calc_1 \ln+x \left(\frac{3^{3/4}}{6} \calc_1+\frac{125}{1728}\ 3^{3/4}\ C+\frac{875}{864}\ 3^{1/4}\right)+\calo(x^2\ln x)\,,
\eqlabel{horha2}
\end{equation}
where $\calc_0$ and $\calc_1$ are the two arbitrary integration constants.
On the other hand, the homogeneous solution to \eqref{eom27} is
\begin{equation}
\ha_2^{hom}(v)=\calc_0^{hom}\ \frac{v^8+9}{v^4}\ +\calc_1^{hom}\left(\left(\frac{v^4}{864}+\frac{1}{96 v^4}\right) \ln\frac{3-v^4}{3+v^4}
+\frac{1}{144}\right)\,,
\eqlabel{a2h}
\end{equation}
leading to asymptotic expansion as $v\to 0_+$
\begin{equation}
\ha_2^{hom}(v)={9\ \calc_0^{hom}}\ \frac{1}{v^4}+\calc_0^{hom}\ v^4-\frac{\calc_1^{hom}}{972}\ v^8+\calo(v^{16})\,,
\eqlabel{assa2}
\end{equation}
and 
\begin{equation}
\ha_2^{hom}(x)=6 \calc_0^{hom}+\frac{1}{144} \calc_1^{hom} 
\ln 2-\frac{1}{576} \calc_1^{hom} \ln 3+\frac{1}{144} \calc_1^{hom}+\frac{1}{144} \calc_1^{hom} \ln x
+\calo(x)\,,
\eqlabel{homa2hor}
\end{equation}
as $x\to 0_+$.
Thus, given \eqref{horha2}, we can take 
\begin{equation}
\calc_1^{hom}=-144 \calc_1\,,
\eqlabel{homtune}
\end{equation}
and tune\footnote{This is always possible since $\cals_{(2,6)}=\calo(v^8)$, see \eqref{source26}.}  $\calc_0^{hom}$
so that 
$$
\ha_2^{v}+\ha_2^{hom}(v)
$$
both have only a normalizable mode as $v\to 0_+$, and be finite (along with its derivatives) as $x\to 0_+$

We now have all the necessary ingredients to determine $\he_1$ from the nonsingularity of  
$\cali^{[2]W}_2(v)$. We find that $\cali^{[2]W}_2(v)$  is finite as $v\to 3^{1/4}_-$, provided
\begin{equation}
\he_1=\frac{3^{1/4}\sqrt{2}}{432} \left(7182-15\ \delta_1+2^{5/2}\ 3^{3/4}\ \delta_2\right)\,.
\eqlabel{eta01}
\end{equation}
Notice that while $\he_0$ \eqref{eta0} is determined unambiguously from the nonsingularity condition of the 
background geometry, the absence of singularities is not a powerful enough constraint to fix $\he_1$. 
As we already mentioned, this fact will not preclude us from computing a definite ratio of shear viscosity
to the entropy density.  

\subsubsection{$\calr_{\mu\nu}\calr^{\mu\nu}$}
Analysis of the square of the Ricci tensor can be performed in the same way as for the Riemann tensor square.
We find
\begin{equation}
\calr_{\mu\nu}\calr^{\mu\nu}={\rm finite}-\ga\ \frac{1920\ \hal_2(v)}{\t^{4/3}}\,,\qquad v\to 3^{1/4}_-\,,
\eqlabel{ricci}
\end{equation}
were we explicitly indicated dependence on order-2 fields.
We argued above that $\hal_2(v)$ can be chosen to be finite as $v\to 3^{1/4}_-$; this would guarantee the 
nonsingularity of $\calr_{\mu\nu}\calr^{\mu\nu}$ to orders $\calo(\t^{-4/3})$ and $\calo(\ga)$.

\subsubsection{Higher order curvature invariants}
As in \cite{bbhj} we denote
\begin{equation}
\calr^{[2^n]}\ _{\mu\nu\r\l}\equiv \calr^{[2^{n-1}]}\ _{\mu_1\nu_1\mu\nu}\cdot \calr^{[2^{n-1}]}\ ^{\mu_1\nu_1}\ _{\r\l}\,,
\eqlabel{riemann}
\end{equation}
where
\begin{equation}
\calr^{[0]}\ _{\mu\nu\r\l}\equiv \calr_{\mu\nu\r\l}\,.
\eqlabel{n0}
\end{equation}
We further define higher curvature invariants $\cali^{[2^n]}$, generalizing \eqref{i2def}:
\begin{equation}
\begin{split}
\cali^{[2^n]}&\equiv \calr^{[2^{n-1}]}\ _{\mu\nu\r\l}\calr^{[2^{n-1}]}\ ^{\mu\nu\r\l}\\
&=\biggl(\cali^{[2^n]SUGRA}_0(v)+\ga\ \cali^{[2^n]W}_0(v)\biggr)+\frac{1}{\t^{2/3}} \biggl(\cali^{[2^n]SUGRA}_1(v)+
\ga\ \cali^{[2^n]W}_1(v)\biggr)\\
&+\frac{1}{\t^{4/3}} \biggl(\cali^{[2^n]SUGRA}_2(v)+\ga\ \cali^{[2^n]W}_2(v)\biggr)+\calo(\t^{-2})+\calo(\ga^2)\,.
\end{split}
\eqlabel{indef}
\end{equation}

Given complexity of analysis, we checked at order $\calo(\t^{-4/3})$ only\footnote{We studied more general higher curvature invariants 
at lower orders in the late proper time expansion.}  nonsingularity of $\cali^{[4]}$.
Keeping explicit only fields at order-2, we find
\begin{equation}
\begin{split}
\cali^{[4]W}_2(v)=&-\frac{55296 v^5 (v^4-3)}{(3+v^4)^7} \biggl(v^{16}-4 v^{12}+198 v^8-36 v^4+81\biggr)\ f_2'(v)\\
&+\frac{663552 v^9 (v^4-3)^3}{(3+v^4)^7}\ \hal_2'(v) 
-\frac{1536}{(3+v^4)^8} \biggl(5 v^{32}+120 v^{28}+6876 v^{24}\\
&+33480 v^{20}-72738 v^{16}+301320 v^{12}+556956 v^8+87480 v^4\\
&+32805\biggr)\
 \hal_2(v)+\dd\cali_4(v)\,,
\end{split}
\eqlabel{i4w}
\end{equation}
No new analysis of order-2 fields are necessary here to verified that \eqref{i4w} is nonsingular precisely for \eqref{eta01}.

\section{Shear viscosity for the Bjorken flow of $\caln=4$ SYM plasma}
In previous section we analytically evaluated $\a'$-corrected supergravity background 
dual to a Bjorken flow of $\caln=4$ SYM plasma at finite coupling to order $\calo(\t^{-2/3})$ 
in the late proper time expansion.  We can now extract the boundary energy density $\e(\t)$ from the one-point correlation function 
of the boundary stress energy tensor using the 
$\a'$-corrected holographic renormalization developed in \cite{bal}. We confirmed that the  final expression for the energy density can 
be evaluated as in the supergravity approximation \cite{j1,j2,bbhj}:
\begin{equation}
\e(\t)=-\frac{N^2}{2\pi^2}\ \lim_{v\to 0}\frac{2 a(v,\t)}{v^4\t^{4/3}}\,.
\eqlabel{etau}
\end{equation} 
Using \eqref{defform}, \eqref{da0}, \eqref{da1}, \eqref{etaex}, \eqref{eta0} and \eqref{eta01} we find
\begin{equation}
\e(\t)=\frac{N^2(6+576\ \ga +\ga\ \dd_1)}{12\pi^2}\ \frac{1}{\t^{4/3}} -\frac{N^2\ 2^{1/2}\ 3^{1/4}\ (1566 \ga+8+\gamma \delta_1)}{48\pi^2}\
\frac{1}{\t^2}+\calo(\t^{-8/3})\,.   
\eqlabel{etres}
\end{equation}
Notice that even though $\eta_1$ depends on an arbitrary integration constant $\dd_2$, see \eqref{eta01}, 
such a dependence disappears in \eqref{etres}.

The string theory result \eqref{etres} should now be interpreted within Muller-Israel-Stewart theory \cite{m,is}. 
For the Bjorken flow of the $\caln=4$ SYM plasma we expect \cite{bbhj}
\begin{equation}
\e^{gauge}(\t)=\frac 38\ \pi^2 N^2\Lambda^4\ (1+15\gamma)\ \frac{1}{\t^{4/3}}-\pi^2 N^2\Lambda^3 A\ (1+15\gamma)\ \frac{1}{\t^2}+\calo(\t^{-8/3})\,,
\eqlabel{egauge}
\end{equation}
where $\Lambda$ is an arbitrary scale, related to the initial energy density of the expanding plasma, and 
\begin{equation}
A=\frac{\eta}{s}\,,
\eqlabel{defa}
\end{equation}
is the ratio of shear viscosity to entropy density. 
Notice that in \eqref{egauge} we have to use $\caln=4$ equation of state at finite 't Hooft coupling\footnote{We 
would like to thank Romuald Janik for pointing this out.}, see \cite{gkt}. 
Matching the leading terms in the late time expansion in \eqref{etres} and \eqref{egauge}
we find
\begin{equation}
\Lambda^4=\frac{4}{3\pi^4}\biggl(1+81\ga+\frac 16\ \ga\ \dd_1\biggr)+\calo(\ga^2)\,,
\eqlabel{deflambda}
\end{equation}
which implies that 
\begin{equation}
A=\frac{1}{4\pi}\biggl(1+120\ \ga\biggr)+\calo(\ga^2)\,,
\eqlabel{final}
\end{equation}
leading to the result quoted in \eqref{floweta}. 
Notice that the remaining  arbitrary integration constant $\dd_1$ got ``absorbed'' into a definition of an arbitrary scale $\Lambda$.
The physical observable, \ie, the ratio of shear viscosity to  entropy density \eqref{final} is evaluated unambiguously.

\section{Conclusion}
In this paper we evaluated finite 't Hooft coupling correction to strongly coupled $\caln=4$ SYM plasma undergoing 1+1-dimensional Bjorken expansion.
Such finite coupling correction corresponds to string theory $\a'$ corrections in the supergravity dual to the boost invariant 
plasma expansion proposed by Janik and collaborators \cite{j1,j2,j3}. We extracted the ratio of  shear viscosity to entropy density 
at large, but finite, 't Hooft coupling. Our result is in conflict with analysis of shear viscosity of strongly coupled SYM
plasma extracted from the equilibrium two-point correlation functions of the boundary stress energy tensor \cite{sh1,sh2}.

Though we do not currently understand the source of the discrepancy, there are two potential outcomes.
\nxt First, the nonsingularity approach of \cite{j1,j2,j3} is simply inconsistent --- it was shown 
in \cite{bbhj} that it is definitely inconsistent within the supergravity approximation at high
orders in the late proper time expansion  ( though it appears to be well-defined  in the first two orders necessary 
to compute shear viscosity of $\caln=4$ plasma in the supergravity approximation ).     
\nxt The discrepancy might be due to using incorrect effective string theory action \eqref{aaa}, as a dual to $\caln=4$
dynamics at large, but finite, 't Hooft coupling. Specifically, the full set of $\a'$ corrections to RR fluxes is unknown.
It is conceivable that proper inclusion of $\a'$ corrections to the RR fluxes would reconcile two results (as it must be).
A starting point for such analysis could be the $\a'$ effective action proposed in \cite{hss}.

We believe resolution of the puzzle presented here  deserves further study.

\section*{Acknowledgments}
My research at Perimeter Institute is supported in part by the Government
of Canada through NSERC and by the Province of Ontario through MRI.
I gratefully acknowledges further support by an NSERC Discovery
grant and support through the Early Researcher Award program by the
Province of Ontario.

\appendix

\section{Appendix: Source terms for eqs.~\eqref{eom11}-\eqref{eom17}}
\begin{equation}
\begin{split}
\cals_{(1,1)}=-\frac{72 \he v^6 \dd_1}{(3+v^4)^4}+\frac{1119744 (20 v^{16}-340 v^{12}+1010 v^8-555 v^4-9) \he v^6}{(3+v^4)^9}
\end{split}
\eqlabel{source11}
\end{equation}
\begin{equation}
\begin{split}
\cals_{(1,2)}=&-\frac{72 (v^{16}+54 v^8+72 v^4+81) \he v^6 \dd_1}{(v^8-9)^4}-\frac{373248\he v^6}{(3+v^4)^9 (v^4-3)^4}\times
 \biggl(
113319 v^{16}\\
&-55647 v^4+159732 v^8-180090 v^{12}-37251 v^{20}+2187-532 v^{28}+12 v^{32}\\
&+6750 v^{24}\biggr)
\end{split}
\eqlabel{source12}
\end{equation}
\begin{equation}
\begin{split}
\cals_{(1,3)}=&-\frac{144 (v^{16}+6 v^{12}+54 v^8+18 v^4+81) \he v^6 \dd_1}{(v^8-9)^4}-\frac{746496\he v^6}{(3+v^4)^9 (v^4-3)^4}\times
 \biggl(24 v^{32}\\
&-604 v^{28}+6840 v^{24}-37017 v^{20}+113751 v^{16}-179118 v^{12}+139806 v^8\\
&-30861 v^4+2187\biggr)
\end{split}
\eqlabel{source13}
\end{equation}
\begin{equation}
\begin{split}
\cals_{(1,4)}=&-\frac{12 v^6 \he (v^4+15) (v^8+9) \dd_1}{(3+v^4)^4 (v^4-3)^2}+\frac{186624 
 \he v^6}{(v^4-3)^2 (3+v^4)^9}\times \biggl(84 v^{24}-2312 v^{20}\\
&+15663 v^{16}-50778 v^{12}+74304 v^8
-37098 v^4-1215\biggr)
\end{split}
\eqlabel{source14}
\end{equation}
\begin{equation}
\begin{split}
\cals_{(1,5)}=&\frac{24 v^7 \he \dd_1}{(v^8-9) (v^4-2 v^2+3) (v^4+2 v^2+3)}
\\
&-\frac{373248  v^7 \he}{(v^4+2 v^2+3) (v^4-2 v^2+3) (v^4-3) (3+v^4)^8}\times \biggl(40 v^{20}-520 v^{16}\\
&+1455 v^{12}-1269 v^8-189 v^4-81\biggr)
\end{split}
\eqlabel{source15}
\end{equation}
\begin{equation}
\begin{split}
\cals_{(1,6)}=&\frac{259200 (71 v^8+639-456 v^4) v^{14} \he}{(3+v^4)^9 (v^4-3)}
\end{split}
\eqlabel{source16}
\end{equation}
\begin{equation}
\begin{split}
\cals_{(1,7)}=&\frac{288 \he v^{10} \b_2}{(v^8-9)^3}-\frac{44789760 v^{14} \he (v^8-9 v^4+27) (v^8-3 v^4+3)}{(v^4-3)^3 (3+v^4)^9}
\end{split}
\eqlabel{source17}
\end{equation}

\section{Appendix: Source terms for eqs.~\eqref{eom21}-\eqref{eom27}}
We find it convenient to express the source terms at this order implicitly in the background warp factors $\{a_2,b_2,c_2\}$, 
while using equations of motion for the supergravity background to eliminate higher than second derivatives\footnote{The highest derivative is 
the fourth.}. For convenience, we present these equations here \cite{j2,bbhj}
\begin{equation}
\begin{split}
a_2''=&-\frac{9+5 v^4}{v (v^4-3)} a_2'-\frac{16 v^3}{v^8-9} b_2'-\frac{32 v^3}{v^8-9} c_2'
+\frac{4(2 v^4+9) v^4}{3(v^4-3) (v^8-9)}\\
&-\frac{432 v^6 \he^2 (v^{16}+24 v^{12}+30 v^8+216 v^4+81)}
{(v^8-9)^4}\,,
\end{split}
\eqlabel{a2eq}
\end{equation}
\begin{equation}
\begin{split}
b_2''=&-\frac{5 v^8-8 v^4+27}{v (v^8-9)} b_2'+\frac{16 v^3}{v^8-9} c_2'-\frac{432 v^6 \he^2 (v^{12}+11 v^8+63 v^4-27)}{(v^8-9)^3 (3+v^4)}\\
&- \frac{4(2 v^8-9 v^4-63) v^4}{3(v^4-3)^2 (v^8-9)}\,,
\end{split}
\eqlabel{b2eq}
\end{equation}
\begin{equation}
\begin{split}
c_2''=&\frac{8 v^3}{v^8-9} b_2'-\frac{5 v^8-16 v^4+27}{v (v^8-9)} c_2'-\frac{432 v^6 \he^2(v^{12}-13 v^8-9 v^4-27)}{(3+v^4) (v^8-9)^3} \\
&+\frac{4(v^8-9 v^4-18) v^4}{3(v^4-3)^2 (v^8-9)}\,.
\end{split}
\eqlabel{c2eq}
\end{equation}
Of course, \eqref{order2} solves \eqref{a2eq}-\eqref{c2eq}.

\begin{equation}
\begin{split}
\cals_{(2,1)}=&\frac{4320 v^7 (119 v^{16}-837 v^{12}+4776 v^8-7533 v^4+9639)}{(v^4-3) (3+v^4)^7} a_2'
+\biggl(\frac{2 v^3 \dd_1}{(3+v^4)^2}\\
&-\frac{864  v^3}{(3+v^4)^7 (v^4-3)^3}(125 v^{28}-15725 v^{24}+145095 v^{20}-669570 v^{16}\\
&+1303911 v^{12}-1275669 v^8+96957 v^4+8748)
\biggr) (b_2'+2 c_2')\\
&+\biggl(-\frac{(2 v^8+3 v^4+9) v^4}{(v^4-3)^3 (3+v^4)^2}+\frac{432 v^6 (2 v^{12}-21 v^8+36 v^4-81) \he^2}{(3+v^4)^5 (v^4-3)^3}\biggr) \dd_1
\\
&-\frac{72 (v^8-2 v^4+9) v^6 \he \dd_2}{(v^4-3)^2 (3+v^4)^4}
+\frac{373248 v^6\he^2}{(v^4-3)^6 (3+v^4)^{10}}\times \biggl(-87058152 v^{16}\\
&-22973922 v^{24}+61231383 v^{20}-39438171 v^8+92195901 v^{12}+7794468 v^4\\
&+177147+6804783 v^{28}-1074783 v^{32}+126421 v^{36}+132 v^{44}-6007 v^{40}\biggr)\\
& -\frac{288 v^4}{(3+v^4)^7 (v^4-3)^6}\times \biggl(v^{40}+2386 v^{36}+5073 v^{32}+50304 v^{28}+2354787 v^{24}\\
&-3096360 v^{20}+26563545 v^{16}
-11763144 v^{12}+26305236 v^8-118098 v^4\\
&-118098\biggr)
\end{split}
\eqlabel{source21}
\end{equation}
\begin{equation}
\begin{split}
\cals_{(2,2)}=&-\biggl(\frac{2 v^3 \dd_1}{(v^4-3)^2}+\frac{288 v^3 }
{(3+v^4)^7 (v^4-3)^2}(535 v^{24}-5190 v^{20}+34755 v^{16}-92925 v^{12}\\
&+160542 v^8-109593 v^4+8748)
\biggr) a_2'
+\frac{288 v^7}{(3+v^4)^7 (v^4-3)^3} \biggl(473 v^{24}-18797 v^{20}\\
&+177291 v^{16}-787422 v^{12}+1595619 v^8-1522557 v^4+344817\biggr) b_2'\\
&+\biggl(-\frac{24 v^7 \dd_1}{(v^8-9)^2}+\frac{576 v^7}{(3+v^4)^7 (v^4-3)^3} (41 v^{24}-14909 v^{20}+162387 v^{16}\\
&-751134 v^{12}
+1471203 v^8-1184301 v^4+47385)
\biggr) c_2'\\
&+\biggl(\frac{v^4 (4 v^{16}-6 v^{12}-171 v^8-648 v^4-567)}{9(v^4-3)^4 (3+v^4)^2}
+\frac{144 v^6  \he^2}{(v^8-9)^5}(2 v^{20}-99 v^{16}-36 v^{12}\\
&-1458 v^8-2430 v^4-2187)\biggr) \dd_1
-\frac{24 \he v^6 (3 v^{16}-8 v^{12}+90 v^8+216 v^4+243) \dd_2}{(v^8-9)^4}\\
&-\frac{124416 v^6\he^2}{(v^4-3)^6 (3+v^4)^{10}}\times \biggl(-86541048 v^{16}-23414778 v^{24}+65133639 v^{20}\\
&-43348527 v^8+96948981 v^{12}+13463172 v^4
-531441+7098399 v^{28}\\
&-958419 v^{32}+104165 v^{36}+60 v^{44}-2587 v^{40}\biggr) +\frac{32v^4}{(3+v^4)^7 (v^4-3)^6}\times \biggl(4 v^{44}\\
&+21 v^{40}+11406 v^{36}-3843 v^{32}+51372 v^{28}
+14188743 v^{24}-27797580 v^{20}\\
&+159619653 v^{16}-94425912 v^{12}+135077868 v^8-5314410 v^4-2480058\biggr)
\end{split}
\eqlabel{source22}
\end{equation}
\begin{equation}
\begin{split}
\cals_{(2,3)}=&-\biggl(\frac{4 v^3 \dd_1}{(v^4-3)^2}+\frac{576 v^3}{(3+v^4)^7 (v^4-3)^2} (535 v^{24}-5190 v^{20}+34755 v^{16}-92925 v^{12}\\
&+160542 v^8
-109593 v^4+8748)\biggr) a_2'
+\biggl(-\frac{24 v^7 \dd_1}{(v^8-9)^2}+\frac{576 v^7}{(3+v^4)^7 (v^4-3)^3}\\
&\times (41 v^{24}-14909 v^{20}+162387 v^{16}-751134 v^{12}+1471203 v^8
-1184301 v^4\\
&+47385)\biggr) b_2'+\biggl(-\frac{24 v^7 \dd_1}{(v^4-3)^2 (3+v^4)^2}+\frac{1152 v^7}{(3+v^4)^7 (v^4-3)^3}
 (257 v^{24}-16853 v^{20}\\
&+169839 v^{16}-769278 v^{12}+1533411 v^8-1353429 v^4+196101)\biggr) c_2'
\\
&+\biggl(\frac{2v^4 (v^{16}+12 v^{12}+99 v^8+162 v^4+162)}{9(v^4-3)^4 (3+v^4)^2}
-\frac{288 v^6 \he^2}{(v^8-9)^5}(4 v^{20}+99 v^{16}+252 v^{12}\\
&+1458 v^8+2187)\biggr) \dd_1
-\frac{48 v^6 \he (3 v^{16}+16 v^{12}+90 v^8+243) \dd_2}{(v^8-9)^4}\\
&-\frac{248832 v^6\he^2}{(v^4-3)^6 (3+v^4)^{10}}\times \biggl(-87964056 v^{16}-24187842 v^{24}+60165747 v^{20}\\
&-30830139 v^8+80345277 v^{12}+3542940 v^4-531441+6773211 v^{28}\\
&-1132371 v^{32}+126641 v^{36}+168 v^{44}-6439 v^{40}\biggr)+\frac{64v^4}{ (3+v^4)^7 (v^4-3)^6} \times \biggl(v^{44}\\
&+12 v^{40}+3873 v^{36}
+57582 v^{32}-870975 v^{28}+10771434 v^{24}-39529377 v^{20}\\
&+117880758 v^{16}-124357194 v^{12}+97483338 v^8+1417176 v^4+708588\biggr)
\end{split}
\eqlabel{source23}
\end{equation}
\begin{equation}
\begin{split}
\cals_{(2,4)}=&-\biggl(\frac{v^3 (v^4-3) \dd_1}{4(3+v^4)^2}
+ \frac{9v^3}{2(3+v^4)^7 (v^4-3)} (16 v^{28}+192 v^{24}-57909 v^{20}+492612 v^{16}\\
&-3870990 v^{12}+5165316 v^8-13920741 v^4+69984)\biggr) a_2'
\\
&+\biggl(-\frac{v^3 (v^{12}-33 v^8+27 v^4-675) \dd_1}{12(v^8-9)^2}
-\frac{3v^3}{2 (3+v^4)^7 (v^4-3)^3}(16 v^{36}+288 v^{32}\\
&+54939 v^{28}-6400314 v^{24}+62833977 v^{20}-333665244 v^{16}+710087229 v^{12}\\
&-937477962 v^8
+81104895 v^4+15746400)
\biggr) b_2'
\\
&+\biggl(\frac{v^3 (v^{12}+3 v^8-27 v^4+459) \dd_1}{3(v^8-9)^2}
+\frac{6 v^3}{(3+v^4)^7 (v^4-3)^3} (16 v^{36}+144 v^{32}\\
&-11157 v^{28}+3087459 v^{24}-31309542 v^{20}+166984254 v^{16}-355695705 v^{12}\\
&+476607807 v^8
-55436076 v^4-10707552)
\biggr) c_2'
+\biggl(\frac{72 (v^4+1) v^2 \dd_1}{(v^8-9)^2}
\\
&-\frac{144 v^2}{(3+v^4)^7 (v^4-3)^2} (16 v^{28}+240 v^{24}+1371 v^{20}-48555 v^{16}+48465 v^{12}\\
&-196425 v^8-186624 v^4-69984)\biggr) (b_2+2 c_2)
+\biggl(\frac{6 \he^2 v^2 (2 v^8-9 v^4-27) \dd_1}{ (v^8-9)^2}
\\
&-\frac{324 \he^2 v^2}{(3+v^4)^7 (v^4-3)^2} (8 v^{28}+120 v^{24}+623 v^{20}+17985 v^{16}-21555 v^{12}+36315 v^8\\
&+13608 v^4-5832)
-\frac{4 \he v^2 \dd_2}{v^8-9}\biggr) 
\ln\frac{3-v^4}{3+v^4}
+\biggl(-\frac{(v^4-9) v^8}{(v^4-3)^3 (3+v^4)^2}\\
&+\frac{12 v^6\he^2}{(3+v^4)^5 (v^4-3)^3} (-1917 v^8+12 v^{16}-81 v^{12}-2511 v^4-8019+4 v^{20})\biggr) \dd_1\\
&-\frac{16 v^6 \he (v^{12}+12 v^8+27 v^4+108) \dd_2}{(v^4-3)^2 (3+v^4)^4}
+\frac{216 v^6\he^2}{(v^4-3)^6 (3+v^4)^{10}}\\
&\times \biggl(100639406448 v^{20}+177616911942 v^{12}+19659065472 v^4+931084632\\
&-84883350843 v^8
-148217241528 v^{16}+165520368 v^{36}-1463056992 v^{32}\\
&+10018391364 v^{28}-34345317222 v^{24}+98934 v^{44}-6831063 v^{40}+312 v^48\\
&+32 v^{52}\biggr)
+\frac{2 v^2}{(v^4-3)^6 (3+v^4)^7}\times
\biggl(-17006112 v^6-289315557 v^{26}\\
&-3796250733 v^{18}+74994660 v^{22}-4935413835 v^{10}+931858830 v^{14}\\
&-13966290 v^{30}-476019 v^{34}-207648 v^{38}
-144 v^{42}\biggr)
\end{split}
\eqlabel{source24}
\end{equation}
\begin{equation}
\begin{split}
\cals_{(2,5)}=&\biggl(\frac{(v^4-3) v^4 \dd_1}{6(v^4+2 v^2+3) (v^4-2 v^2+3)}
+\frac{48 v^4}{(v^4-2 v^2+3) (v^4+2 v^2+3) (3+v^4)^6} (v^{28}\\
&+12 v^{24}-9 v^{20}-6948 v^{16}+15435 v^{12}-65934 v^8-10935 v^4-4374)\biggr) a_2'
\\
&+\biggl( \frac{v^4 (v^4+1) \dd_1}{6(v^4-2 v^2+3) (v^4+2 v^2+3)}
\\
&+\frac{48 v^4}{(v^4-2 v^2+3) (v^4+2 v^2+3) (3+v^4)^6 (v^4-3)^2} (v^{36}+14 v^{32}+24 v^{28}\\
&+1782 v^{24}-136782 v^{20}+553644 v^{16}
-1220832 v^{12}+161838 v^8-2187 v^4\\
&+13122)
\biggr) (b_2'+2 c_2')
-\biggl( \frac{(v^4-6) v^9}{9(v^4-3)^2 (v^4+2 v^2+3) (v^4-2 v^2+3)}\\
&+\frac{72 v^7 (v^8-18) \he^2}{(v^4-2 v^2+3) (v^4+2 v^2+3) (v^8-9)^2}\biggr) \dd_1
\\
&+\frac{24 \he v^7 \dd_2}{(v^4-2 v^2+3) (v^4+2 v^2+3) (v^8-9)}\\
&-\frac{20736 v^7\he^2}{(v^4-3)^5 (3+v^4)^9 (v^4+2 v^2+3) (v^4-2 v^2+3)}\times
 \biggl(1062882+18680058 v^{24}\\
&+59248017 v^{16}+7853517 v^8+1417176 v^4-33907248 v^{12}-36367380 v^{20}\\
&+1161 v^{40}-46404 v^{36}+732348 v^{32}-4038876 v^{28}
+v^{48}
+12 v^{44}\biggr)\\
&-\frac{32v^9}{(v^4-2 v^2+3) (v^4+2 v^2+3) (3+v^4)^6 (v^4-3)^5}\times \biggl(
v^{40}+6 v^{36}-117 v^{32}\\
&-12276 v^{28}-112041 v^{24}+345492 v^{20}-7735095 v^{16}+6890508 v^{12}-21065184 v^8\\
&+3346110 v^4+236196\biggr)
\end{split}
\eqlabel{source25}
\end{equation}
\begin{equation}
\begin{split}
\cals_{(2,6)}=&\frac{180 v^7 (163 v^{16}-42 v^{12}-2106 v^8-378 v^4+13203)}{(3+v^4)^7 (v^4-3)} a_2'
\\
&+\frac{180 v^7}{(3+v^4)^7 (v^4-3)^3} (163 v^{24}-2556 v^{20}+34557 v^{16}-165528 v^{12}+311013 v^8\\
&-207036 v^4+118827) (b_2'+2 c_2')+\frac{207360 v^{14}\he^2}{(v^4-3)^4 (3+v^4)^{10}} \times\biggl(-166212 v^{12}\\
&-393660 v^4+166941+451413 v^8+229 v^{24}
-4860 v^20+50157 v^{16}\biggr)\\
&-\frac{240v^8}{(3+v^4)^7 (v^4-3)^4}\times \biggl(173 v^{24}+183 v^{20}\\
&+18162 v^{16}-2160 v^{12}+195615 v^8+18225 v^4+7290\biggr)
\end{split}
\eqlabel{source26}
\end{equation}
\begin{equation}
\begin{split}
\cals_{(2,7)}=&\biggl(\frac{v^3 \b_2}{v^8-9}-\frac{466560 v^{11} (2 v^8-11 v^4+18)}{(3+v^4)^7 (v^4-3)}\biggr) a_2'
+\biggl(\frac{v^3 \b_2}{v^8-9}\\
&+\frac{155520 v^{11} (2 v^8-63 v^4+18)}{(3+v^4)^7 (v^4-3)}\biggr) (b_2'+2 c_2')+\biggl(\frac{12 v^4 (v^4+1)}{(v^4-3)^4 (3+v^4)}
\\
&+\frac{10368 v^{10} \he^2}{ (v^8-9)^4}\biggr) \b_2+\frac{288 \he v^{10} \b_4}{ (v^8-9)^3}-\frac{13436928v^{14}\he^2}{(v^4-3)^4 (3+v^4)^{10}} \times
\biggl(11 v^{24}-192 v^{20}\\
&+1853 v^{16}-5832 v^{12}+16677 v^8-15552 v^4+8019\biggr)+\frac{124416v^{16}}{(3+v^4)^7 (v^4-3)^4}\\
&\times \biggl(10 v^{12}+33 v^8+63 v^4+79\biggr) 
\end{split}
\eqlabel{source27}
\end{equation}

\end{document}